# NATURE OF SHORT HIGH AMPLITUDE PULSES IN A PERIODIC DISSIPATIVE LAMINATE METAMATERIAL

Pedro Franco Navarro[1], David J. Benson[2], Vitali F. Nesterenko[1, 3]

[1]Department of Mechanical and Aerospace Engineering, University of California, San Diego, La Jolla CA 92093-0411 USA

[2]Department of Structural Engineering, University of California, San Diego, La Jolla CA 92093-0085 USA

[3]Materials Science and Engineering Program, University of California, San Diego, La Jolla CA 92093-0418 USA

**Abstract.** We study the evolution of high amplitude stress pulses in periodic dissipative laminates taking into account the nonlinear constitutive equations of the components and their dissipative behavior. Aluminum-Tungsten laminate was taken as an example due to the large difference in acoustic impedances of aluminum and tungsten, the significant nonlinearity of aluminum constitutive equation at the investigated range of stresses, and its possible practical applications. Laminates with different cell size, which controls internal time scale, impacted by the plates with different thicknesses, determining the incoming pulse duration, were investigated. It has been observed that the ratio of the duration of the incoming pulse to the internal characteristic time determines the nature of the high amplitude dissipative propagating waves – the oscillatory shock-like profile, the train of localized pulses or a single localized pulse. These localized quasistationary waves resemble solitary waves even in the presence of dissipation: the similar pulses emerged from different initial conditions, indicating that they are inherent properties of the corresponding laminates, their characteristic length scale is determined by the mesostructural scale and the stress amplitude, and they mostly recover their shapes after collision with phase shift and a linear relationship exists between their speed and amplitude. A theoretical description approximating the shape, length scale and speed of these high amplitude dissipative pulses was proposed based on the KdV type equation with a dispersive term dictated by the mesostructure. The nonlinear term is derived from nonlinear constitutive equations of Aluminum and Tungsten, which were extracted from their Hugoniot curves.

## I. INTRODUCTION

The response of materials to high amplitude shock wave loading is usually analyzed based on the assumption that shock wave has reached a steady state which is assumed to happen when the shock wave propagates a distance of a few shock front widths [1]. This assumption allows the use of the conservation laws across the shock front, resulting in the Rankine-Hugoniot equations connecting the states in front and behind the stationary shock wave, if duration of pulse is long enough [1-3].

The case of laminated materials presents a challenge for such an approach, and the first studies to characterize the behavior of these materials in the acoustic realm, including their dispersive properties due to periodic mesostructure, can be found in [4-6]. For example, in [6] the authors propose a one-dimensional lattice model that includes geometrical dispersion. Their model agrees with the results of ultrasonic experiments. The behavior of Al/W metamaterials (W fibers placed in Al matrix) under impact by thick Al plate, generating a semi-infinite loading pulse to avoid the formation of release waves on the back of the flyer plate, was investigated in [7]. The experimental results related to shock rise time show good agreement with the numerical modeling based on the lattice model and nonlinear elastic-plastic response. The structuring of the wave front occurred near the impact surface. The formation of steady shock wave profile in laminate materials can take much longer distances because the establishment of steady state behind shock can be delayed due to the longer mechanical and thermal relaxation processes. This is clearly illustrated in periodic systems composed from metal or glass plates with gaps between them where establishment of the steady state behind shock wave happened after about 10 reverberations of waves behind leading part of the pulse [8,9]. In case when laminate materials are loaded by a long duration incoming disturbance (especially for small size of the cell in laminate material) the stationary shock wave can be formed and the final state can still be described in the frame of Hugoniot approach if conditions of thermal and mechanical equilibrium are hold. In this case the Hugoniot curve is not sensitive to the mesostructural properties of material, e.g., cell size.

In some applications a composites (e.g., laminates) are subjected to a very short duration loading pulses, created for example by powerful lasers. But in the case of the short incoming pulse, caused by impact of thin plate with thickness comparable to the thickness of cell in the laminate or contact explosion of thin layer of explosive [9], or by laser excitation [10,11] the stationary shock wave is not formed. In this case the Hugoniot curve is not representative to describe the state of dynamically compressed material and its response can be very sensitive to the pulse length and material mesostructure. For example, at short duration of incoming pulse interesting phenomena like anomalous dependence of leading shock amplitude on the cell size (increase of amplitude with decrease of cell size) were observed in experiments and in numerical calculations [9,12,13,14] being in contradiction with expected based on acoustic approximation [15]. Nonlinearity in stress strain relation was responsible for these deviation from linear elastic behavior.

In the case of relatively short duration of loading pulse a combination of dispersive properties and nonlinearity in layered media give rise to a qualitatively new, solitary like pulses were numerically investigated in [16-22].

The paper [16] considers the evolution of nonlinear elastic waves in a nondissipative layered media excited by the motion of the boundary and explore the effects of impedance mismatch between layers which is a cause for a dispersion. They observed numerically that at the investigated boundary conditions the impedance mismatch coupled with the nonlinearities, introduced by nonlinear stress-strain dependence, resulted in a transformation of incoming pulse into train of localized pulses. The authors called them *stegotons* since it was not clear whether these localized pulse were formally solitons. They, noticed that width of each stegoton is about 10 layers and it is independent of the size of the layers but it depends on the amplitude. Their speed was equal to the effective sound speed for the linearized medium plus a term linearly depending on the amplitude. Their shape was roughly approximated by the $\text{sech}^2$ function of time, both properties are characteristic for the KdV solitons. Two stegotons with different amplitudes were excited at the boundary and the stegoton with the higher amplitude travels faster and eventually overtakes the first one with a smaller amplitude. As in the case of interacting solitary wave the colliding stegotons roughly assumed their initial shape after separation with some phase shift.

In [16] authors considered a comparison between a nonlinear laminate material with one of the layers having very small density and bulk elastic modulus (ratio of densities being equal to $10^{-3}$ and ratio of bulk moduli being equal to $2.5 \cdot 10^{-4}$) with the Toda lattice where particles of similar masses alternate with springs. They selected the unit cell in Toda lattice being equal to the unit cell of the laminate (composed from the two different layers) and masses of particles being equal to the average density from the layered medium. The spring layers had exponential dependence on the strain mimicking forces in Toda lattice with the coefficient in exponent equal to the smallest bulk modulus. This interaction force resulted in a correct effective sound speed in a linearized medium. In this special limited choice of parameters of laminate material authors found that solitary like behavior of laminate can be modeled directly by Toda lattice. Authors also introduce a set of homogenized equations which support the solitary like solutions similar to the direct solutions of the original hyperbolic system.

The paper [17] emphasized that materials microstructure characterized by intrinsic space-scales, e.g. lattice period, size of crystalline grain, thickness of layers in laminates, distance between microcracks is responsible for the effects of dispersion and being combined with nonlinear behavior results in phenomena like solitons. The authors used discrete and continuum approach for the modeling of effects of microstructure and nonlinearity following the Mindlin approach [18]. Depending on the initial and boundary conditions they demonstrated formation of a single solitary like pulse or the train of these pulses. They obtained good agreement with experimental data [19] of the stress history in the laminate polycarbonate (layer thickness 0.39 mm)-stainless steel (0.19 mm) at the distance 3.44 from impacted face using finite-volume algorithm and nonlinear parameter only for the polycarbonate. The simplified boundary conditions in calculations to model plate impact in experiments were given by a step function with the velocity amplitude equal to the velocity of impacted plate 1043 m/s, thickness 2.87 mm. But the amplitude of the particle velocity in experiments on the interface with laminate (having polycarbonate as the first layer) and polycarbonate impactor with velocity 1043 m/s should be equal halve of this value - 521.5 m/s. Also it is not clear what a duration was selected of the used step function (authors probably used rectangular function) as a boundary conditions modeling impact by a plate with finite thickness which provided the good fit to experimental data.

The propagation and head-on collisions of localized initial pulse were numerically investigated in [20] for microstructural materials with different values of microscale nonlinearity parameter. It was found that the interaction between localized waves is not completely elastic. The phase shift observed in such interactions is increased when the amplitudes are dissimilar and the longer the time the phenomena is observed. Over short time intervals and small number of collisions the behavior of these localized pulses was close to the solitonic behavior in all considered cases.

The emergence of two solitary trains from the localized initial condition was numerically investigated using a Boussinesq-type equation [21] for nonlinear microstructured solids also following Mindlin approach [18]. On the long time scale they observed establishment of the stationary amplitudes of each emerging localized pulse in the train with linear dependence of their speed on the amplitude. The speed and amplitudes of localized pulses were traced after their multiple collisions. They display main characteristics of classical solitons though their interaction is not fully elastic especially at the low amplitudes. Thus authors suggested to use term "quasi-solitons" to distinguish them from "pure solitons" characteristic for highly idealized nonlinear dispersive systems.

The propagation of nonlinear longitudinal strain waves in the case of uniaxial strain sate in a layered material was studied using a macroscopic Boussinesq type wave equation that has been derived in a long wave approximation through a high-order asymptotic homogenization method [22]. The derived coefficients in this equation are related to the linear and nonlinear elastic moduli, densities and volume fractions of the components. The dispersion relation in the linearized Boussinesq type wave equation was close to the exact dispersion relation obtained using the Floquet-Bloch approach for two laminate materials (a low contrast steel-aluminium and high contrast steel-carbon/epoxy) at ratio of cell size to characteristic wave length less than 0.4, better agreement was obtained for high contrast laminate. Authors also obtained parameters (speed and width) of the localized supersonic bell-shaped compression solitary wave as functions of nonlinear properties of materials and amplitude of solitary wave. In case of physically linear elastic materials the geometric nonlinearity in combination with dispersion support only a tension solitary waves.

The authors of [22] numerically investigated the non-stationary dynamic processes integrating macroscopic Boussinesq type wave equation focusing on the evolution of nonlinear waves from different initial excitations. The initial conditions were taken as a rectangular profile with different non-dimensional width $\delta$. In case of a relatively narrow pulse of compression ($\delta=5$) the nonlinear excitation with given amplitude resulted into two solitary wave propagating symmetrically in opposite directions with a scattering backward radiation. It is important that at given composite parameters and amplitude of excitation the formation of steady-shape solitary waves required a spatial interval a three orders of magnitude larger than cell size. A wider pulse of initial compression ($\delta=20$) with the same amplitude resulted in a train of four solitary waves with noticeable amplitudes. The properties of these solitary waves emerging from initial conditions were similar to the stationary solitary wave solutions of the Boussinesq type wave equation. The narrow pulse of tension ($\delta=5$) with the same amplitude did not evolve into a steady-shape wave producing instead a delocalization of the initial excitation.

In this paper we investigate the nature of relatively short pulses in the laminate Al-W (with the cell size equal 1, 2 and 4 mm) generated by impact of 8 mm and 2 mm impactors with velocity 2800 m/s which created realistic initial and boundary conditions reproducible in physical experiments. The dynamic response of materials at this level of stresses is characterized by significant viscoplastic dissipation which was not considered in the previous papers related to the propagation of solitary like pulses (stegotons) in laminates [16-22]. The dissipation can suppress the formation of these solitary like pulses because the disturbance must travel a relatively long distance from the point of excitation before this localized pule was formed, e.g., in [22] the formation of steady-shape solitary like waves in a non-dissipative laminate required a spatial interval a three orders of magnitude larger than cell size. We introduced viscous type dissipation in addition to plastic behavior and selected the corresponding value of the viscosity to match the dissipation behind stationary shock fronts in Al and W matching temperatures in calculations to the values predicted based on Hugoniot relations. The selected values of viscosity ensures the shock front thickness being much smaller than the characteristic length of layers in the laminates.

We investigated the characteristic length travelled by the pulses when single solitary like pulses (or their trains) were formed under realistic, experimentally reproducible plate impact conditions (at various thicknesses of impactor) and their subsequent rate of attenuation in the presence of dissipation. It was found that the ratio of impactor mass to the cell mass determines either a train of solitary like waves or a single traveling pulse were formed. This behavior is consistent with the observed in strongly nonlinear granular systems [9].

We also investigated the head-on collisions of solitary-like pulses impacting laminates from both sides. It was proven that after interaction the most prominent change is a phase shift which agrees with the expected behavior of solitary like waves. It was also observed that interactions are not completely elastic exhibiting some changes in shape of pulses after their interaction due to the dissipation processes and also due to the fact that solitary like pulses had speed lower than sound speed in the layers.

We introduced phenomenological model combining the dispersive properties of laminates, due to their periodic structure, and physical nonlinear properties of materials based on their shock adiabat (Hugoniot) instead of using third elastic moduli as in [22]. Hugoniot curve is representing inherent nonlinear behavior of materials in a stationary shock wave at very broad range of pressures. The combination of dispersive and nonlinear properties resulted in a KdV solitary wave. Based on this approach equations relating speed and widths of these solitary like pulses to the geometrical and physical properties of materials in laminates were presented. Shape of the KdV solitary wave provided satisfactory fit to the shape of the localized pulses observed in numerical calculations. The speeds of KdV solitary waves also were close to the speed of localized pulses in the laminate at similar maximum pressures.

## II. SIMULATIONS

We apply LS-DYNA's simulations to analyze propagation of high amplitude dissipative nonlinear waves generated by plate impact in a 1-D laminate material. Fig 1 represents the geometry used in our numerical modeling, where $L_c$ is the total length of the laminate composite, $L_i$ is the length of the impactor and $a$ and $b$ are the respective thicknesses of the aluminum (Al) and tungsten (W) layers in the periodic composite material. In this paper, all layers have the same thickness ($a=b$). For the purposes of this study we selected $L_c$ = 280 mm for all cases. The duration of incoming pulses was determined by the thickness of impactor equal to 2 and 8 mm to produce a stress waves with different durations and investigate their properties.

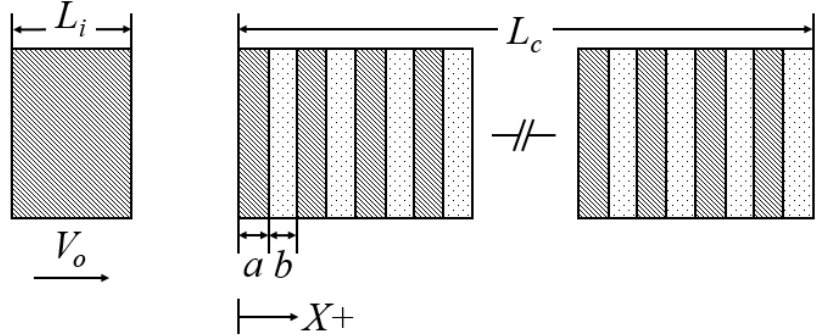

FIG. 1. Geometry of the laminate.

The impactor in the current study was an aluminum plate having high initial velocity of 2800 m/s in the X+ direction. In all cases, the layers of the laminate are perfectly bonded and the end of the composite is free. It is expected that the bonding between layers does not affect the compression shock wave structure and the properties of the final state behind shock because we are considering only compression waves similar to [12-14].

The material model used to characterize the behavior is the well-known Steinberg-Guinan model [23, 24] coupled with the Grüneisen's equation of state, this rate independent model is described as follows

$$G = G_0 \left[1 + AP\eta^{-1/3} - B\left(\frac{E-E_c}{C_p} - 300\right)\right], \tag{1}$$

where $G$ and $G_0$ are current and initial shear moduli, $P$ is the pressure, $\eta = V_0/V$, $V$ and $V_0$ are current and initial specific volumes, $A$ is a coefficient in the pressure dependence of the shear modulus, $B$ is the coefficient in the temperature dependence of the shear modulus and $E$ is the total internal energy per unit volume. The cold energy of the system per unit volume $E_c$ is defined as

$$E_c = \int_{V_0}^{V} PdV - 300 C_p \exp\left\{a\left(1 - \frac{1}{\eta}\right)\right\} \eta^{\gamma_0 - a}, \tag{2}$$

where integral in the first term is on the zero Kelvin isotherm.

The melting energy (in the code it is used to check if material reached the melting conditions during the dynamic deformation) is defined as

$$E_m = E_c + C_p T_m, \tag{3}$$

The dependence of melting temperature $T_m$ on specific volume based on modified Lindemann law is expressed as

$$T_m = T_{m0} \exp\left\{2a\left(1 - \frac{1}{\eta}\right)\right\} \eta^{2(\gamma_0 - a - 1/3)}, \tag{4}$$

where $\gamma_0$ is the original Grüneisen gamma, $a$ is the coefficient of the volume dependence of the Grüneisen's gamma.

The yield strength ($Y$) of the material including dependence on pressure and effects of strain hardening and thermal softening is given by

$$Y = \frac{Y_0}{G_0}\left[1 + \beta(\epsilon_i + \epsilon_p)\right]^n \left[1 + AP\eta^{-1/3} - B\left(\frac{E - E_c}{C_p} - 300\right)\right], \tag{5}$$

where

$$Y_0\left[1 + \beta(\epsilon_i + \epsilon_p)\right]^n \leq Y_{max}.$$

In the Eq. (5) $Y_0$ is the initial yield strength and $Y_{max}$ is the value of saturated yield strength, $\epsilon_i$ is the initial plastic strain, $\epsilon_p$ is the equivalent plastic strain added under dynamic loading, and $\beta$ and $n$ are work-hardening parameters.

Based on the Grüneisen's equation of state the pressure can be expressed as:

$$P = \frac{\rho_0 C_0^2 \mu \left[1+\left(1-\frac{\gamma_0}{2}\right)\mu\right]}{1-(S_1-1)\mu} + \gamma_0 E, \tag{7}$$

$$\mu = \frac{\rho}{\rho_0} - 1, \tag{8}$$

where $\rho_0$, $\rho$ refer to the initial and density in deformed state.

The following materials parameters for Al and W used in calculations are presented on Table 1, they were taken from [24].

TABLE 1. Material Properties

| Parameters | Aluminum | Tungsten |
|---|---|---|
| $G_0$ (GPa) | 27.6 | 160 |
| $Y_0$ (GPa) | 0.29 | 2.2 |
| $Y_{max}$ (GPa) | 0.76 | 4 |
| $\beta$ | 125 | 24 |
| $n$ | 0.1 | 0.19 |
| $A$ (GPa$^{-1}$) | 652 | 9380 |
| $B$ (K$^{-1}$) | 6.16·10$^{-4}$ | 1.38·10$^{-4}$ |
| $T_{m0}$ (K) | 1220 | 4520 |
| $C_p$ (J/kg·K) | 287.67 | 43 |
| $\gamma_0$ | 1.97 | 1.67 |
| $a$ | 1.5 | 1.3 |
| $\rho_0$ (kg/m$^3$) | 2785 | 19300 |
| $C_0$ (m/s) | 5328 | 4030 |
| $S_1$ | 1.338 | 1.237 |

The mesh size in all cases was equal $1 \times 10^{-5}$ m = 0.01 mm and selected artificial viscosity resulted in a shock width in Al, W being at least ten times smaller than the smallest layer thickness in the laminate (0.5 mm). This assures that a steady state is reached behind shock waves when they propagate inside each layer on the initial stage of pulse formation following the impact. Of course it is desirable that the shock width is similar to the one found in experiments [25-27], but the width of the hock front is not important for the parameters of the final state as long as the shock width is significantly smaller than the layer thickness. This ensures that the material reaches Hugoniot states at shock loading in each layer. The Hugoniot states are characteristic for stationary shocks and are independent of the specific mechanisms of dissipation defining resulting shock width. The plastic shock width ($\Delta x$) and the rise time ($\Delta t$) at a shock stress 70 GPa were equal $\Delta t = 3.72 \times 10^{-9}$ and a $\Delta x = 3.38 \times 10^{-5}$ m for Aluminum and $\Delta t = 9.69 \times 10^{-9}$ and $\Delta x = 4.79 \times 10^{-5}$ m for Tungsten. Both of the shock widths are about 10 times smaller than the smallest layer thickness in the laminate (0.5 mm), which assures that Hugoniot states behind the shock are reached inside each of the layers.

We validated results of LS-DYNA calculations with selected material properties based on comparison of the final thermodynamic states behind shock waves with available Hugoniot data. The comparison demonstrated that the used model correctly predicts the stress and specific volume in the simulation of single shock in individual materials. Even more important is that the simulations correctly predict the temperatures on the Hugoniots and also the temperatures after unloading for Al and W. This numerical approach was used to investigate the nature of relatively short, high amplitude stress pulses in the Al-W laminate.

## III. RESULTS OF NUMERICAL CALCULATIONS

The nature of the short stress pulses of different durations excited by the impact of Al plate with thicknesses 2 and 8 mm propagating in the laminate materials with different cell sizes at various distances from the impact side are presented below. The main focus is on the nature of the pulses generated at various conditions of loading and to discover nature of propagating pulses and scaling that might be generated by interplay between duration of impact controlled by the thickness of impactor and cell in the laminate materials. The main characteristic time scaling is determined by the ratio ($t_r$) of the duration of incoming pulse and duration of propagation time through the cell approximated by

$$t_r = \frac{2d_{imp}}{C_{imp}} / \frac{d_{lam}}{C_{lam}}, \qquad (9)$$

where $d_{imp}$ and $d_{lam}$ are the size of the impactor and size of the cell in the laminate respectively, $c_{imp}$ and $c_{lam}$ refer to the sound speed of the impactor and the laminate. For the sound speed in the laminate, we use the definition for the average sound speed in a laminate found in [16, 28]. This definition is later used in the theoretical approach developed in this paper. The final expression for the time ratio is:

$$t_r = \frac{2d_{imp}}{C_{imp}} \sqrt{\frac{d_{Al}^2}{C_{Al}^2} + \frac{d_W^2}{C_W^2} + \left(\frac{Z_{Al}}{Z_W} + \frac{Z_W}{Z_{Al}}\right)\left(\frac{d_{Al}d_W}{C_{Al}C_W}\right)}, \qquad (10)$$

where $d_{Al}$ and $d_W$ refer to the Al and W layer sizes of the laminate, $Z_{Al}$ and $Z_W$ ($Z_i = \rho_i C_i$) represent the acoustic impedance of each layer, finally $C_{Al}$ and $C_W$ are the corresponding sound speeds of each layer. For all our studied cases $C_{imp} = C_{Al}$ but $d_{imp}$ has values of either 2 or 8 mm. From here on, the laminated materials are referred as 2+2, 1+1 and 0.5+0.5 laminates, corresponding to 4 mm, 2 mm and 1 mm cells. All the laminates were loaded by impact of Al plate at velocity 2800 m/s and FWHM refers to full width at half maximum.

### A. 2+2 Al/W Laminate, impact by 8 mm Al plate

The impact loading of 2+2 laminate by Al impactor with the thickness 8 mm corresponds to the value of $t_r$ equal 2.5. In these calculations real materials properties presented in the Table 1 were used. Figs. 2 (a)-(d) shows that these conditions of impact did not result in a steady state shock wave in the laminate because the duration of incoming pulse was too short.

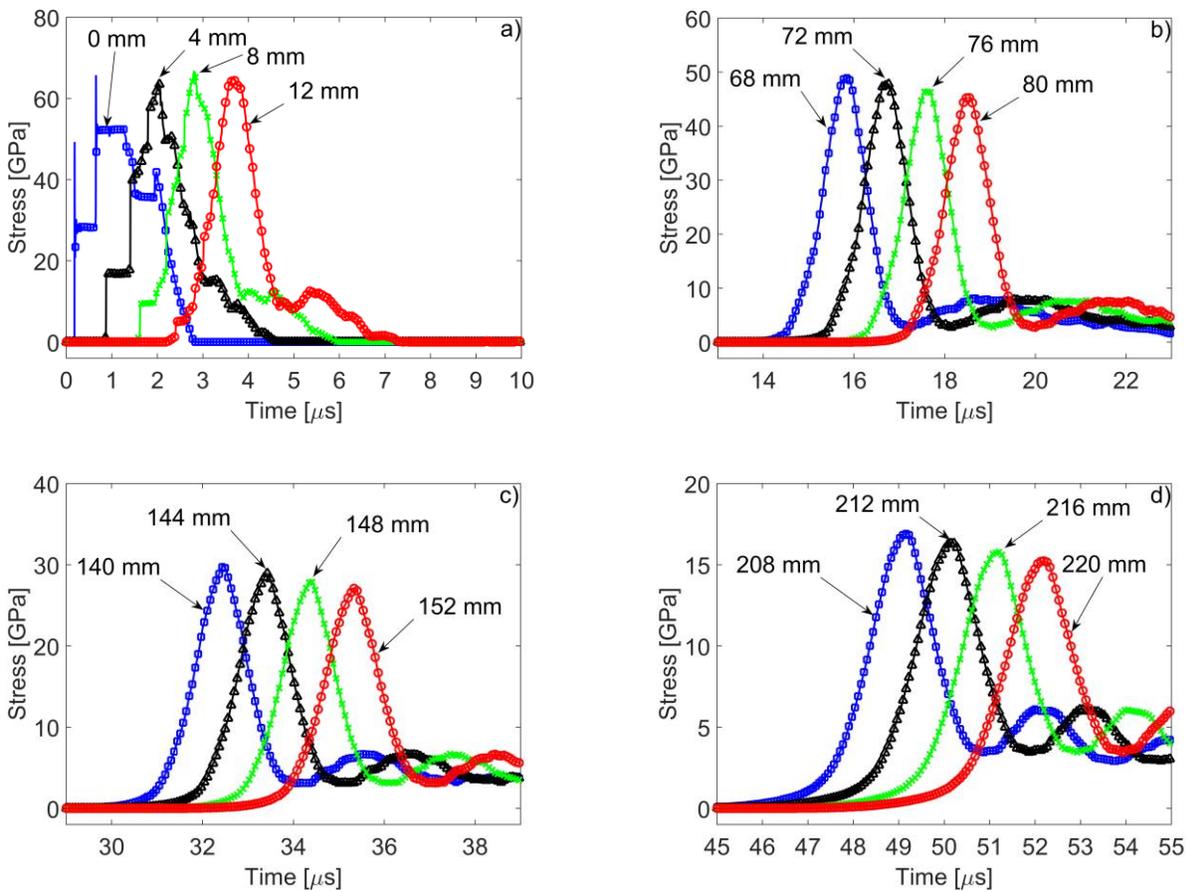

FIG. 2. (Color online) Stress pulse evolution in 2+2 laminate, data corresponds to the interfaces of Al/W layers at different depths (a) 0, 4, 8, and 12 mm (b) 68, 72, 76, and 80 mm (c) 140,

144, 148, and 152 mm (d) 208, 212, 216, and 220 mm. The pulse was generated by the impact of 8 mm Al plate at velocity of 2800 m/s.

Shock waves in each of the layers are clearly distinguishable at depths up to 8 mm from the impacted end. We observe the formation of qualitatively new pulse (not a shock wave) mainly formed at the distance 12 mm from the impacted end (Fig. 2(a). It has a width close to only 3 cell sizes with a small amplitude tail. It is clear from Figs. 2(a)-(d) that formation of the main pulse is caused by multiple reflection of leading stress pulses on Al-W interfaces mostly balanced by nonlinearities of the Al and W behavior. The tail is due to the dissipative properties of materials.

The origin of dispersion is apparently due to multiple shock wave reflections from interfaces clearly seen in Figs. 2 (a). The main pulse propagates with decreasing amplitude being equal 65 GPa at the depth 12 mm and about 15 GPa at depth 220 mm. This decrease of amplitude is accompanied by an increase of the main pulse space width (FWHM increased about 40%). At larger depths the main pulse is accompanied by oscillatory tail and at the relatively short distances, e.g., from 208 to 220 mm the whole pulse can be considered as quasistationary with similar profile but decreasing amplitude (Fig. 2(d)).

It is very important to remark that observed main localized pulse with characteristic length comparable to the cell size of laminate was formed in dissipative media. The rate of dissipation in our calculations matches real losses of energy unlike in previous papers where no dissipation was taken into account [16-18, 20, 21, 22]. It is essential to point out that this "real" dissipation does not prevent formation of prime localized wave in our conditions of loading, and it only results in its attenuation.

We can see that the mesostructure and nonlinear properties of laminate resulted in dramatic transformation of initial shock wave into localized pulse with different path of loading in comparison with shock wave. As a result, temperatures corresponding to the maximum stresses are significantly different than at the same stresses in shock wave. For example, shock wave at stress 55 GPA in Al results in temperature equal 1800 K versus temperature 650 K in the localized pulse.

The formation of the localized pulse with smooth front is beneficial for shock protection because it prevents spall on the free surface.

### B. 2+2 Al/W Laminate with artificially small $Y_{max}$, impact by 8 mm Al plate

We now investigate the laminate with same mesostructure but with reduced maximum yield strength $Y_{max}$ of the materials thus causing plastic flow to occur at lower stress levels reducing dissipation. This case helps us to clarify role of plastic deformation on the formation of the pulse. It is possible to observe that close to the impacted side the same dispersion and nonlinearity shapes the leading wave as in the previous case (compare Fig. 2(a). with Fig. 3(a)). But in case of smaller dissipation the leading wave is formed with the practically unloaded state behind separated from the second small amplitude wave due to difference in their speed of propagation. In the case of material with real properties we formed a quasistationary attenuating oscillatory wave profile (Figs. 2(a)-(d)).

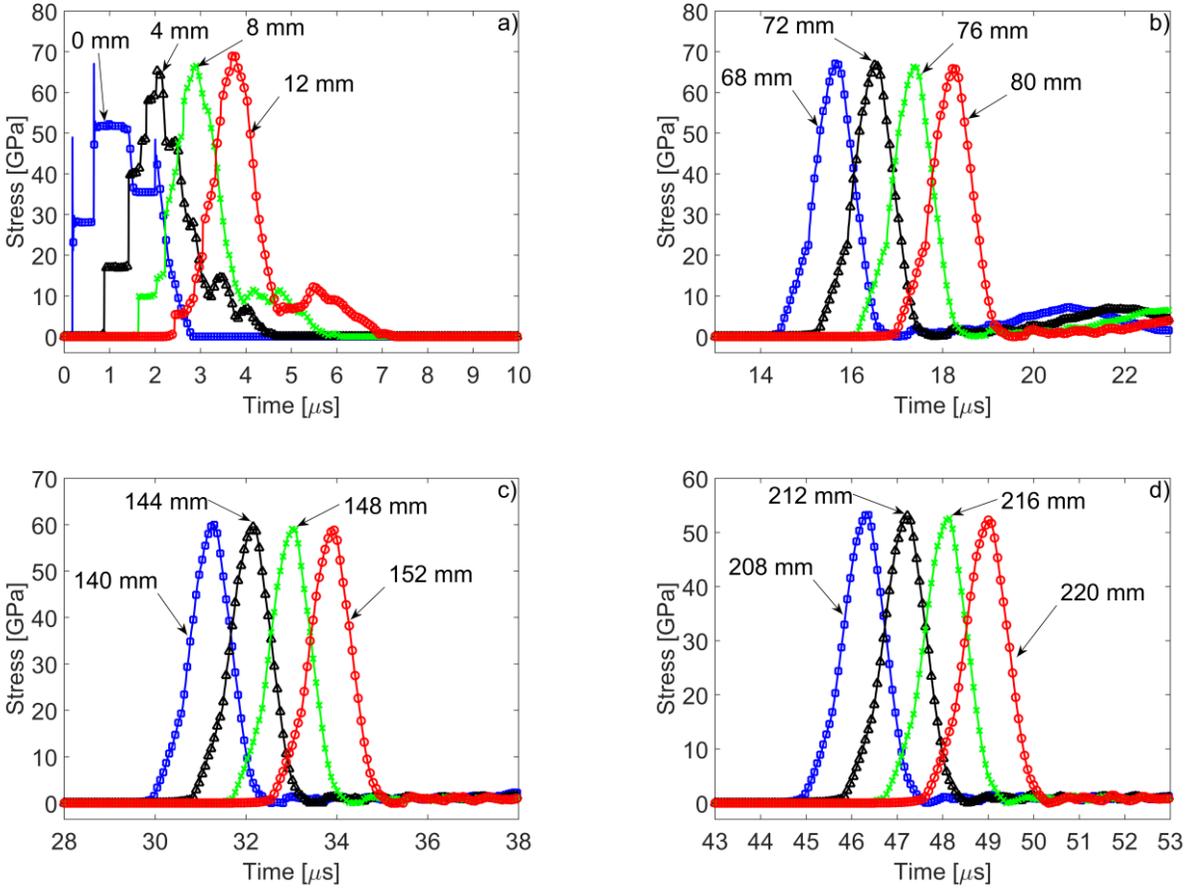

FIG. 3. (Color online) Stress pulse evolution in 2+2 laminate with artificially small $Y_{max}$, data corresponds to the interfaces of Al/W layers at different depths (a) 0, 4, 8, and 12 mm (b) 68, 72, 76, and 80 mm (c) 140, 144, 148, and 152 mm (d) 208, 212, 216, and 220 mm. The pulse was generated by the impact of 8 mm Al plate at velocity of 2800 m/s.

It is interesting to compare pulses in material with real dissipative properties, near the impacted end, pulses with similar amplitude in artificially low dissipative materials and pulses of similar amplitude where laminate behaves elastically. Fig. 5 demonstrates that these pulses at similar maximum amplitudes (taken at different distances from the impacted side, 40 mm for real properties, 156 mm for artificially low dissipation and 80 mm for the elastic case) are almost identical. In the case of the material with artificially low maximum yield strength there is a slight difference in the pulse shape - the front of the wave has a clear kink that does not disappear until the pulse has traveled about 200 mm Fig. 3(d). To understand how these pulses propagate in a pure elastic laminate, we ran calculations with real dissipative properties up to the propagation distance where the quasistationary pulse was formed after travelling the distance 64 mm and after that it transmitted into the pure elastic material with identical mesostructure. The pulse in pure elastic materials propagated initially with relatively small changes (amplitude increase from initial value of 55 GPa to 59 GPa) quickly approaching the steady state after distance of about 11 cells (Figs. 4).

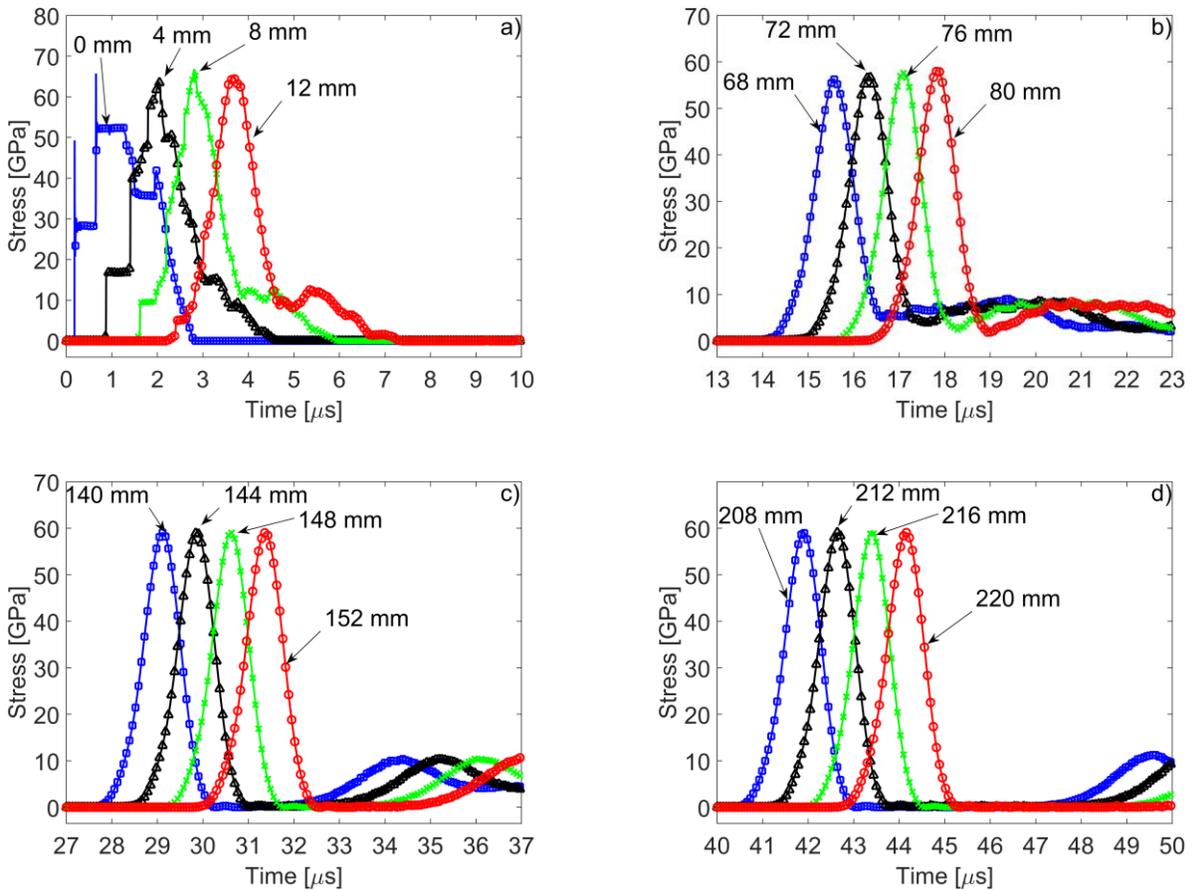

FIG. 4. (Color online) Stress pulse evolution in 2+2 laminate, data corresponds to the interfaces of Al/W layers at different depths (a) 0, 4, 8, and 12 mm (b) 68, 72, 76, and 80 mm (c) 140,

144, 148, and 152 mm (d) 208, 212, 216, and 220 mm. The pulse was generated by the impact of 8 mm Al plate at velocity of 2800 m/s. Initial length of the laminate with real dissipative properties was 64 mm, it was in contact with pure elastic laminate with identical mesostructure.

Thus it is possible to conclude, that the leading pulse in dissipative material even at relatively large distances from the impacted end (Figs. 2, 3, and 3) is dominated by the combination of dispersion and nonlinearity and cell size dictates the spatial dimension of this stress pulse (FWHM of the pulse is equal to 1.17 cell sizes).

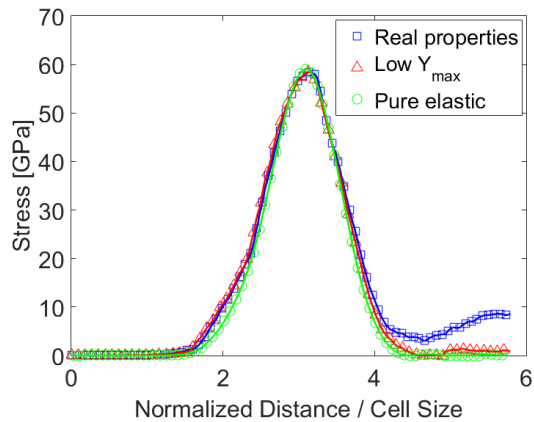

FIG. 5. (Color online) Comparison of wave shape at same stress amplitude of 58 GPa generated by the impact of Al plate with thickness 8 mm for the case of a 2+2 laminate with real dissipative properties, laminate with an artificially small $Y_{max}$, and laminate with elastic behavior after 64 mm.

Despite the similarity of stress pulses (Fig. 5) we may observe different temperatures due to difference in dissipative properties. The time dependence of the stresses and temperatures in the localized pulse propagating in the 2+2 laminates with different dissipative properties is presented in the Fig. 6. It is clear that the amplitude decrease in both cases (less in the material with an artificially small $Y_{max}$) did not significantly change the pulse duration time and in the case of the material with smaller dissipative properties the shape of the pulse was maintained practically the same, with the FWHM change being only about 3%. This agrees with the previous conclusion that the pulse shape is determined mostly by nonlinearity and dispersion with dissipation affecting mostly amplitude.

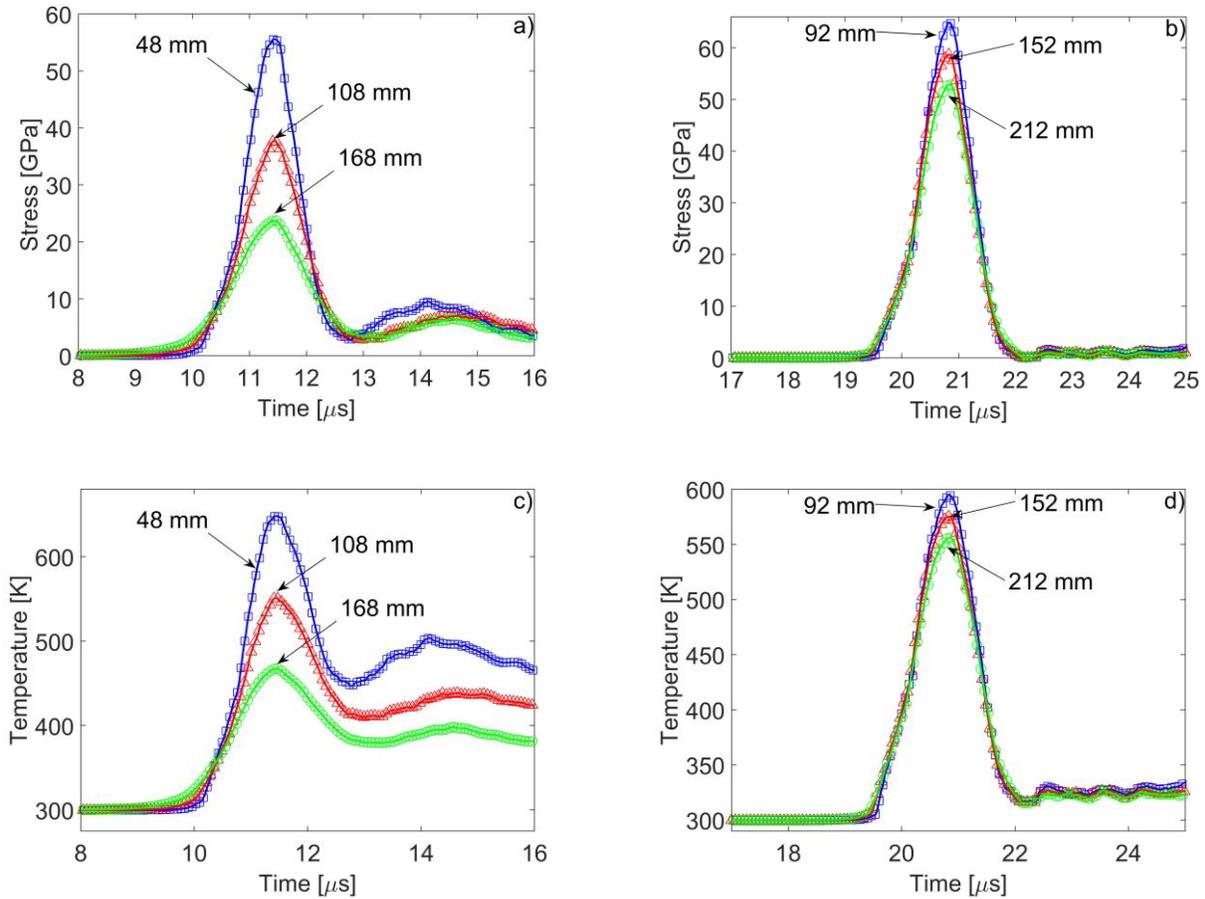

FIG. 6. (Color online) Comparison of attenuating stress traveling waves generated by the impact of Al plate with 8 mm thickness in a 2+2 laminate and their corresponding temperatures in Al layer: (a) and (c) - laminate with real material properties and (b) and (d) – laminate with an artificially small $Y_{max}$.

Though nonlinearity and periodic mesostructure in real materials are able to support the localized pulse with the shape similar to the less dissipative case, the important feature in the former laminate is the significant residual temperature, about half of the maximum temperature (Fig. 6(c)). This behavior emphasizes the difference of this localized wave with classical solitary wave in which material returns to its initial state after being dynamically compressed.

It is very important to emphasize that the maximum temperatures at these waves are significantly smaller than temperatures in the shock waves at the same amplitude. In the material with real dissipative properties at maximum stress 55 GPa in a quasistationary localized wave, increase of Al temperature was 350 K (Fig. 6(c)) which is more than three times smaller than increase of temperature in the shock wave in Al at the same pressure (1154K), the initial temperature was 300 K.

The difference in temperature increase after unloading is also dramatic – after unloading of quasistationary localized pulse with amplitude 55 GPa increase of residual temperature over initial temperature was about 150K, versus increase of residual temperature being 486K after shock loading with the same stress amplitude and subsequent unloading.

If we consider a whole pulse with leading amplitude 55 GPa (including the tail) as a quasistationary shock wave then the estimate of the temperature at the tail with stress 4 GPa (Fig. 6(a)) will be equal 30K (corresponding to shock stress 4 GPa), which much smaller than 170 K in the tail of localized wave (Fig. 6(c)). Thus complex quasistationary pulse with two maxima can't be considered as a stationary shock wave.

These comparisons demonstrate that under considered impact loading temperature estimations based on the Hugoniot of components using dynamic pressures are not adequate. Because the localized pulses provide shockless compression of Al and W it is interesting to compare maximum calculated temperatures with temperature corresponding to isentropic compression. For Al at 55 GPa estimate of the isentropic temperature increase is equal 227 K [29] which is the smaller than the calculated temperature increase 350 K in localized wave (Fig. 6(c)). The estimated temperature in isentropic compression is much closer to the calculated temperature in localized wave than temperature at shock loading at the same pressures. Thus temperatures under isentropic compression give the reasonable lower estimate of the maximum temperatures in localized waves.

### C. 1+1 Al/W Laminate, impact by 8 mm Al plate

It is interesting to determine if the shape and amplitude of the localized pulse observed in previous calculations is scaled with the size of the cell in laminates or it simply determined by incoming pulse. To clarify this point we calculated the wave evolution inside a 1+1 mm laminate generated by the same impact as in case of 2+2 laminate. Thickness of impactor was the same, but time ratio of impactor to the cell was larger than in the previous case. This also allows us to investigate if the number of localized pulses depends on the time ratio. For example, in the case of strongly nonlinear waves in granular materials [9] a single solitary wave was excited when mass of impactor is equal to the mass of the particle and multiple solitary waves were generated at larger mass of impactor. But the nature of propagating pulse can be strongly influenced by dissipative properties of the dispersive media [9, 30, 31].

Impact of 8 mm Al plate on this 1+1 laminate corresponds to time ratio of impactor to cell size equal to 5.1, which is about two times larger than in previous case of 2+2 laminate. Also it is interesting to see if the decrease of the cell size at the same impactor mass results in faster formation of localized pulses and if the corresponding distance for the formation of such pulses is scaled with the cell size.

The evolution of the wave at different depths is presented in Fig. 7. The impact with this impactor/cell time ratio 5.1 demonstrated tendency to a formation of multiple pulses close to impacted end. Nevertheless, the dissipative properties of the material prevent the formation of the train of solitary like waves. Instead an oscillatory triangular profile was observed at larger depths (Figs. 7(c)-(d)).

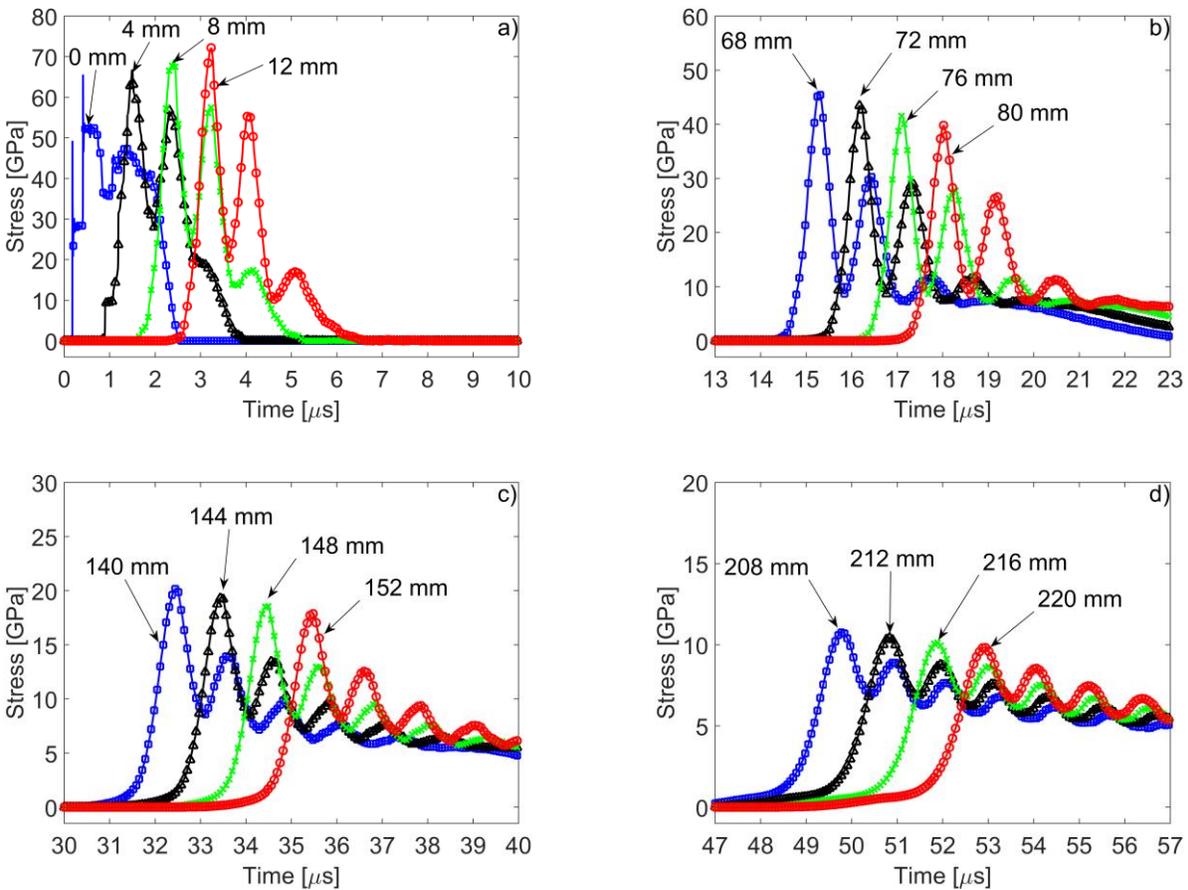

FIG. 7. (Color online) Stress pulse evolution in 1+1 laminate, data corresponds to the interfaces of Al/W layers at different depths (a) 0, 4, 8, and 12 mm (b) 68, 72, 76, and 80 mm (c) 140, 144, 148, and 152 mm (d) 208, 212, 216, and 220 mm. The pulse was generated by the impact of 8 mm Al plate at velocity of 2800 m/s.

There is a difference between maximum amplitudes of stress in these pulses at similar depths 220 mm. For example, amplitude of the leading pulse is equal to 15.2 GPa for 2+2 laminate versus 9.8 GPa for 1+1 laminate, despite the same impact and the same average density of laminates. This difference is apparently due to the difference in their dispersive properties resulting in generation of multiple solitary like wave in the laminate with smaller cell sizes. This demonstrates that amplitudes of the stress pulses can be decreased by decreasing cell size in laminates, which results in multiple solitary like waves with small amplitude instead of a single one with a larger amplitude.

At the depths where the quasistationary pulse is observed (68 to 80 mm), ratio of its characteristic scales to cell size (FWHM/d = 1.25 and (0.1-0.9) $\Delta/d$ = 1.15) are similar to the observed for 2+2 laminate, demonstrating their scaling with the laminate cell size.

### D. 1+1 Al/W Laminate with artificially small $Y_{max}$, impact by 8 mm Al plate

We saw in the previous section that dissipative properties may be responsible for triangular oscillating wave profile preventing formation of train of separate solitary waves. To clarify the role of dissipation we investigated the nature of the wave generated at the same impactor/cell time ratio (5.1), but introducing artificially small yield strength and the reduction of the dissipation. It is clear that reduced dissipative characteristics of the components facilitated the separation of the train of localized pulses (Fig. 8). They travel with different speeds depending on their amplitude and resembling a train of weakly attenuating solitary like waves. This behavior is contrary to the previous case where a quasistationary strongly attenuating triangular oscillatory profile was formed at identical conditions of impact (Fig. 7).

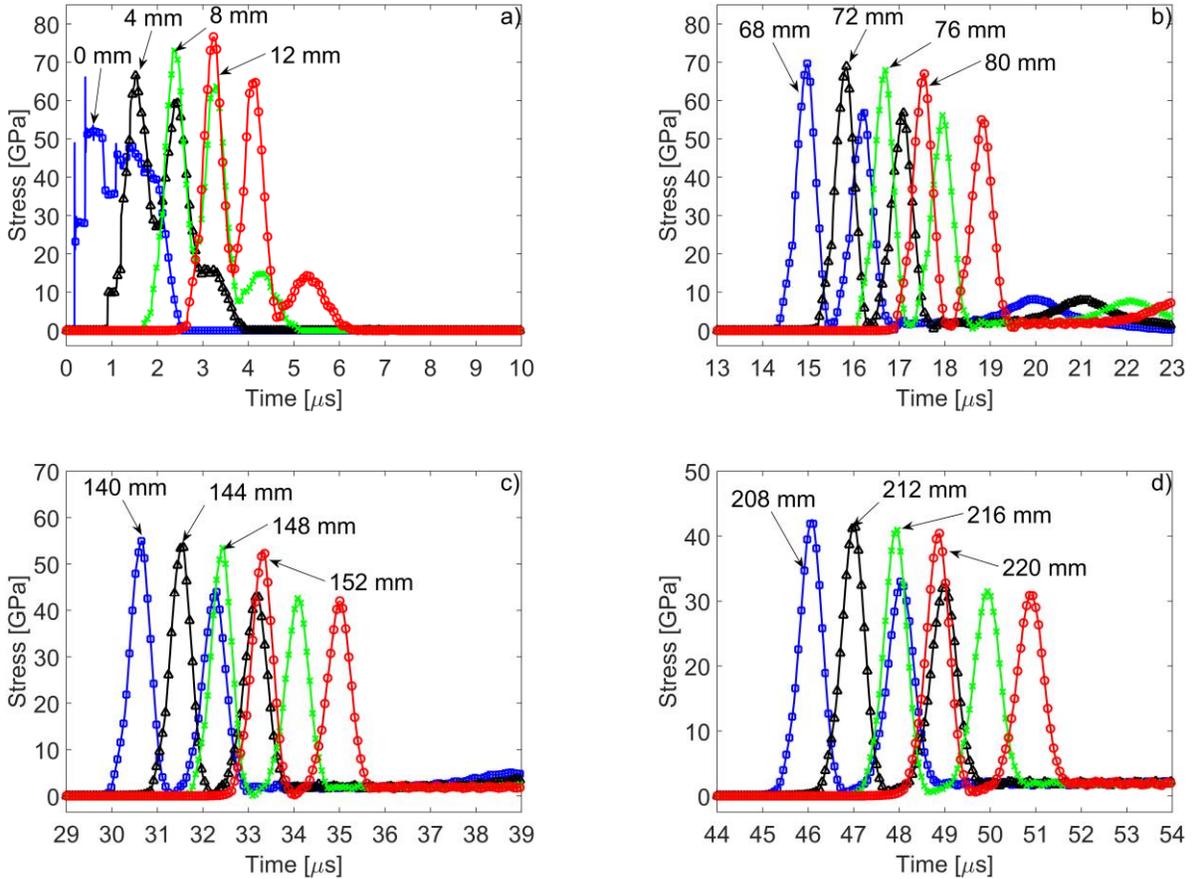

FIG. 8. (Color online) Stress pulse evolution in 1+1 laminate with artificially small $Y_{max}$,, data corresponds to the interfaces of Al/W layers at different depths (a) 0, 4, 8, and 12 mm (b) 68, 72, 76, and 80 mm (c) 140, 144, 148, and 152 mm (d) 208, 212, 216, and 220 mm. The pulse was generated by the impact of 8 mm Al plate at velocity of 2800 m/s.

We consider the formation of a quasistationary solitary like wave when the stress in the wave reached 10% of its maximum on the release part. Following this agreement we observed that its formation occurs fairly fast (at about 26 mm in depth) after travelling 13 cells. The apparently longer distance for establishing a stationary solitary like pulse in this case, in comparison with impact on 2+2 laminate, with the same thickness of impactor is due to the presence of the second wave following the first one.

The FWHM width of localized pulse was equal to 1-1.25 cell size depending on amplitude, which is similar to the width of localized pulse in 2+2, laminate in nondissipative and dissipative cases (Fig. 8).

**E. 0.5+0.5 Al/W Laminate, impact by 8 mm Al plate**

It is interesting to investigate the characteristic scale of localized pulses in relation to cell size and their number and distance at which they are formed in laminate with reduced cell size under the same impact (duration of incoming pulse). The impactor/cell time ratio was increased to 10.2 by reducing the cell size to 1 mm (0.5+0.5 mm layered material) and keeping the duration of incoming pulse the same (thickness of Al impactor 8 mm). The results should be compared with the cases where 2+2 and 1+1 laminates were impacted by the 8 mm thick Al plate at the same velocity (Figs. 2 and 7).

In the case of 0.5+0.5 laminate, an oscillating triangular pulse was formed indicating the tendency to create the train of five localized waves Fig. 9(a), instead of one pulse (in 2+2 laminate, time ratio 2.5) and three pulses (in 1+1 laminate, mass ratio 5.1) in previous cases, also with real dissipative properties of components (Figs. 2 and 7). It should be mentioned that at time ratio equal to 2.5 a single quasi-stationary pulse was formed (Fig. 2) and at mass ratio 5.1 an oscillatory attenuating triangular pulse was observed (Fig. 7). These results prove that there is a strong correlation between shapes of wave profiles generated by the same impact at different time ratios. The time ratio determines the ratio of the characteristic time of incoming load and time scale determined by mesostructured, e.g., time of wave propagation through the cell or time duration of corresponding quasistationary pulses.

In weakly nonlinear and strongly nonlinear discrete materials there is a value of critical viscosity corresponding to the transition from oscillatory stationary shock profile to monotonous shock [32]. It seems that in case of laminates there is a value of yield strength $Y_{max}$ that will prevent splitting of initial pulse into train of solitary waves resulting in oscillatory or monotonous shock like pulse.

Because different wave profiles were formed in laminates with different cell sizes at the same impact, it is interesting to compare the effectiveness of each laminate to decrease the amplitude of leading pulse (it should be mentioned that decrease of cell size may result in the opposing effect and cause an increase of the amplitude of the leading pulse [14] (see also Fig. 2(a), 7(a), 9(a)). The amplitudes of the leading pulses in laminates with different cell sizes, at the same depth 220 mm were equal 15.2 GPa (2+2 laminate), 9.8 GPa (1+1 laminate) and 10.3 GPa (0.5+0.5 laminate). Although there is a small difference between the amplitudes for the 1+1 mm and 0.5+0.5 (with slight increase of amplitude in the latter case), the amplitude in 2+2 layered composite is about 50% larger. This difference in the amplitudes of the leading wave can be explained due to the changes on dispersive properties of the laminate without changing the dissipative and nonlinear properties of the components. These data also demonstrate that amplitude of the leading pulse can't be reduced indefinitely with reduction of the cell size.

Very interesting phenomena of the increase of the duration of the triangular pulse in 0.5+0.5 laminate should be mentioned (compare Figs. 7 and 9). We will see later that this is directly connected to dispersive properties of laminates.

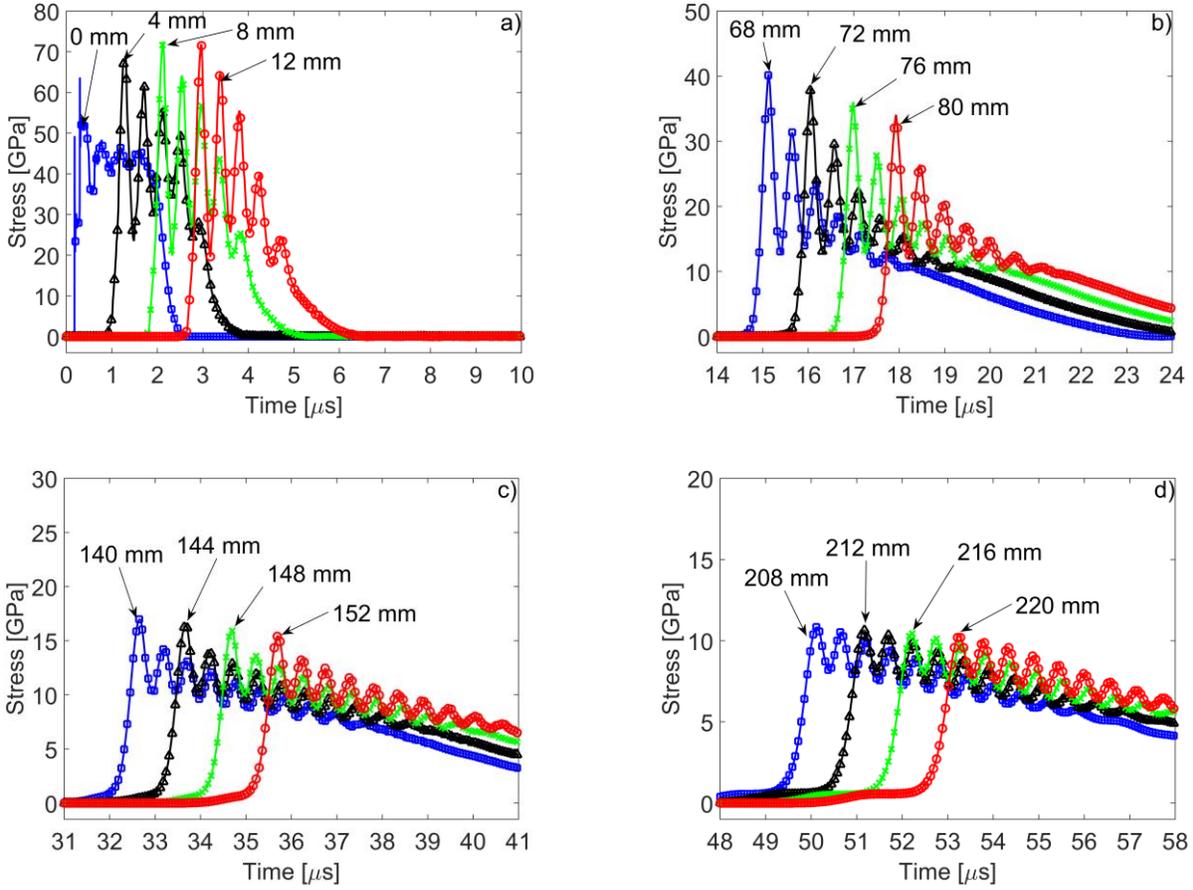

FIG. 9. (Color online) Stress pulse evolution in 0.5+0.5 laminate, data corresponds to the interfaces of Al/W layers at different depths (a) 0, 4, 8, and 12 mm (b) 68, 72, 76, and 80 mm (c) 140, 144, 148, and 152 mm (d) 208, 212, 216, and 220 mm. The pulse was generated by the impact of 8 mm Al plate at velocity of 2800 m/s.

The characteristic (0.1-0.9) space scale of the main leading front of the oscillating triangular wave at large depths is scaled with cell size being about 1.1 cell sizes (Fig. 9(b)). This scaling is similar to the size of the leading front in 1+1 laminate (Fig. 7(b), 1.1 cell size). It is interesting that in this laminate we observe dispersive elastic precursor which lengths is increasing with propagation distance (Fig. 12(d)).

### F. Al/W 2+2 laminate, Al impactor with thickness 2mm

In the case of 2+2 laminate with real material properties, under the impact of an 8 mm Al flyer plate, we observed that a slightly attenuating solitary like pulse was generated from relatively long incoming pulse (Fig. 2). It is interesting to see if a similar solitary like wave could be generated in the same 2+2 laminate from different initial conditions (reduced duration of impact by using 2 mm instead of 8 mm Al flyer plate, in this case the time ratio between the impactor and the cell is 0.6). We could expect that shorter duration impact may create a similar solitary like pulse if the laminate under high amplitude loading behaves as a classical weakly dissipative nonlinear dispersive media.

Fig. 10 (a)-(d) show the evolution of the stress profile as a result of the impact of Al plate with a thickness 2 mm on the 2+2 laminate. It is clear that the localized pulse is also formed at the distance about 24 mm from impacted end (not shown in figures). This process took considerably longer distance than 12 mm in the 2+2 laminate impacted by 8 mm Al plate (compare Fig. 10 with Fig. 2).

Contrary to what we observed in the case were an 8 mm impactor, the amplitude of the formed solitary like pulse was smaller (compare Fig. 2 and 10). This could be due to the smaller linear momentum and energy in case of 2 mm Al striker. But as soon as solitary like pulse was formed (with FWHM equal about 2 cells and 0.1-0.9 ramp size being close to 2 cells also) it was practically identical to the leading part of the same amplitude (10 GPa) pulse in the same laminate formed under impact by 8 mm (Fig. 11). It is important to remark that once the amplitude has been greatly reduced (below 5 GPa at the depths larger than 140 mm) an elastic precursor can be observed and the shape of the pulse doesn't resemble a classic shape of a soliton Fig. 10(c)-(d).

In this case, the solitary like waves with low amplitudes at depths 68 – 80 mm have a FWHM = 2 cell sizes and a 0.1-0.9 ramp size (Δ) equal to 1.9 cell size.

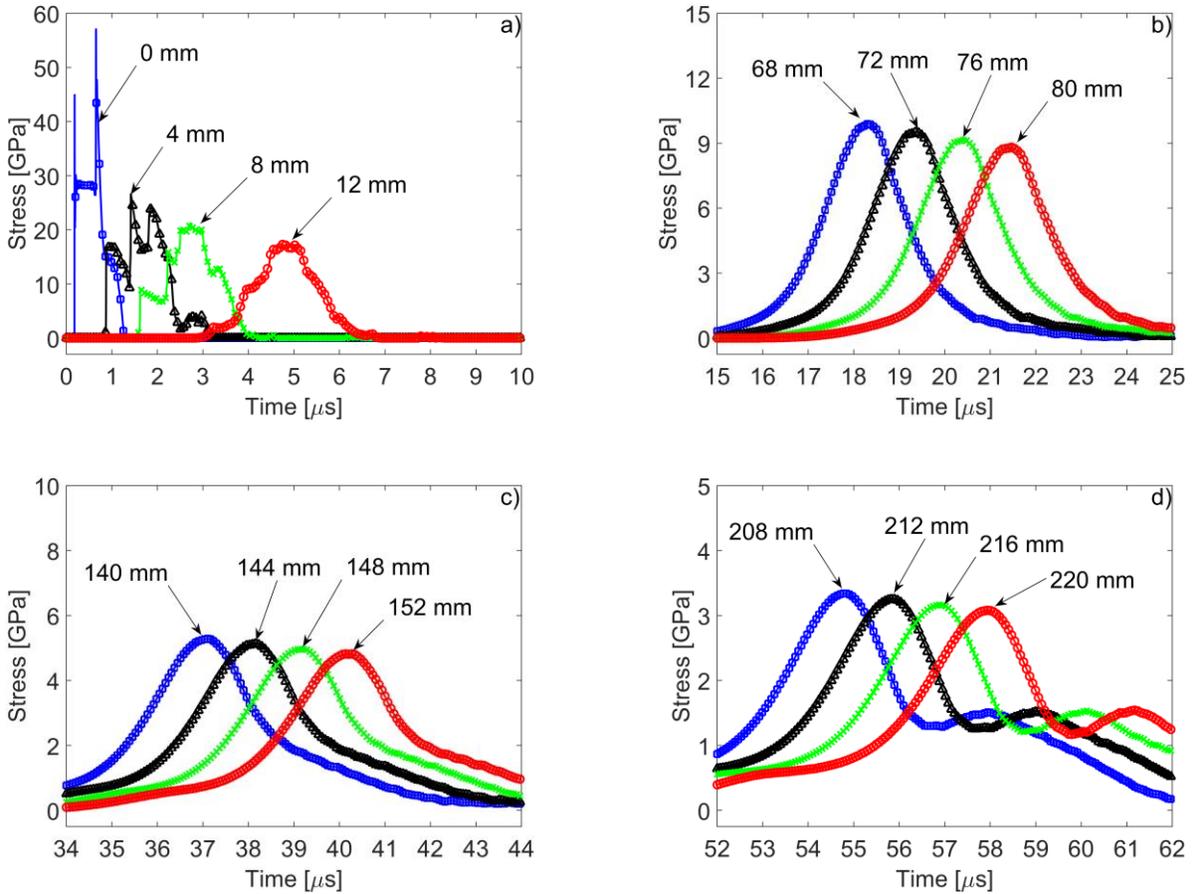

FIG. 10. (Color online) Stress pulse evolution in 2+2 laminate, data corresponds to the interfaces of Al/W layers at different depths (a) 0, 4, 8, and 16 mm (b) 68, 72, 76, and 80 mm (c) 140, 144, 148, and 152 mm (d) 208, 212, 216, and 220 mm. The pulse was generated by the impact of 2 mm Al plate at velocity of 2800 m/s.

It is important to compare localized pulses with similar amplitude propagating in a laminate with the same mesostructure, but excited with different initial conditions (impact by 2 mm Al plate vs. 8 mm Al plate).

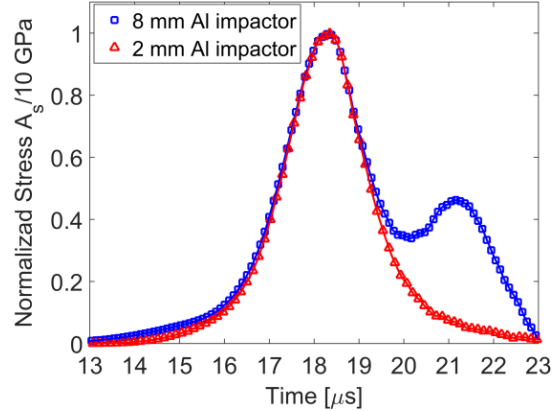

FIG. 11. (Color online) Comparison of shapes of solitary like wave and leading part of the pulse both formed in 2+2 laminates with real dissipative properties excited by impact of Al plate with different thickness, 2 and 8 mm correspondingly.

Figure 11 presents this comparison. It is evident that the resulting attenuating localized pulse has a direct relation to the laminate mesostructure. It is interesting that the rising parts of the localized pulses in the 2+2 laminate are almost identical though they were excited by impact of 8 mm and 2 mm Al plates.

### G. 1+1 Al/W Laminate, impact by 2 mm Al plate

It interesting to study if the localized pulses will be observed in laminates with different cell sizes under identical impact conditions (2 mm Al plate with velocity 2800 m/s) and if their space scale is scaled with the cell size also. The previous case with impactor/cell time ratio of 0.6 (2+2 laminate) resulted in a single solitary like wave (Fig. 10).

Based on what has been observed in 2+2 laminates impacted by 2 mm and 8 mm Al impactors, we can expect the formation of one or more solitary like waves. In the case with 1+1 laminate, impacted by 2 mm Al impactor, the time ratio is 1.3. Fig. 12 demonstrates that this laminate indeed supports the single solitary like wave. The localized pulse is completely formed at a depth of about 8 mm (Fig. 12(a)) compared to 24 mm in the case of 2+2 laminate impacted by the same 2 mm Al plate (compare 12(a) and Fig. 10(a) at the similar depths). It is interesting that decrease of cell size by two times resulted in the three time decrease in the travel distance required to form a solitary like pulse.

It is clear that the shape of the pulse closely resembles a classic solitary wave (bell shape) at a relatively high amplitude of the maximum stress. At lower stress levels (below 5 GPa, Fig. 12(c)-(d)), an elastic precursor can be clearly identified and the shape of the wave does not resemble a classic solitary wave and the pulse propagates as attenuating oscillatory wave. This is consistent with was observed in 2+2 laminate at low stress levels (Fig. 10(c)-(d)). This occurrence can be explained by the unbalance that exists between the nonlinearities and dispersion at low stress levels. Similar to what was observed before, at low amplitudes at depths 68 – 80 mm, the FWHM = 1.8 – 2 cell sizes and $\Delta$ = 1.7-1.9 cell size, which is consistent to the scaling phenomena observed in past cases impacted by 2 or 8 mm Al plate.

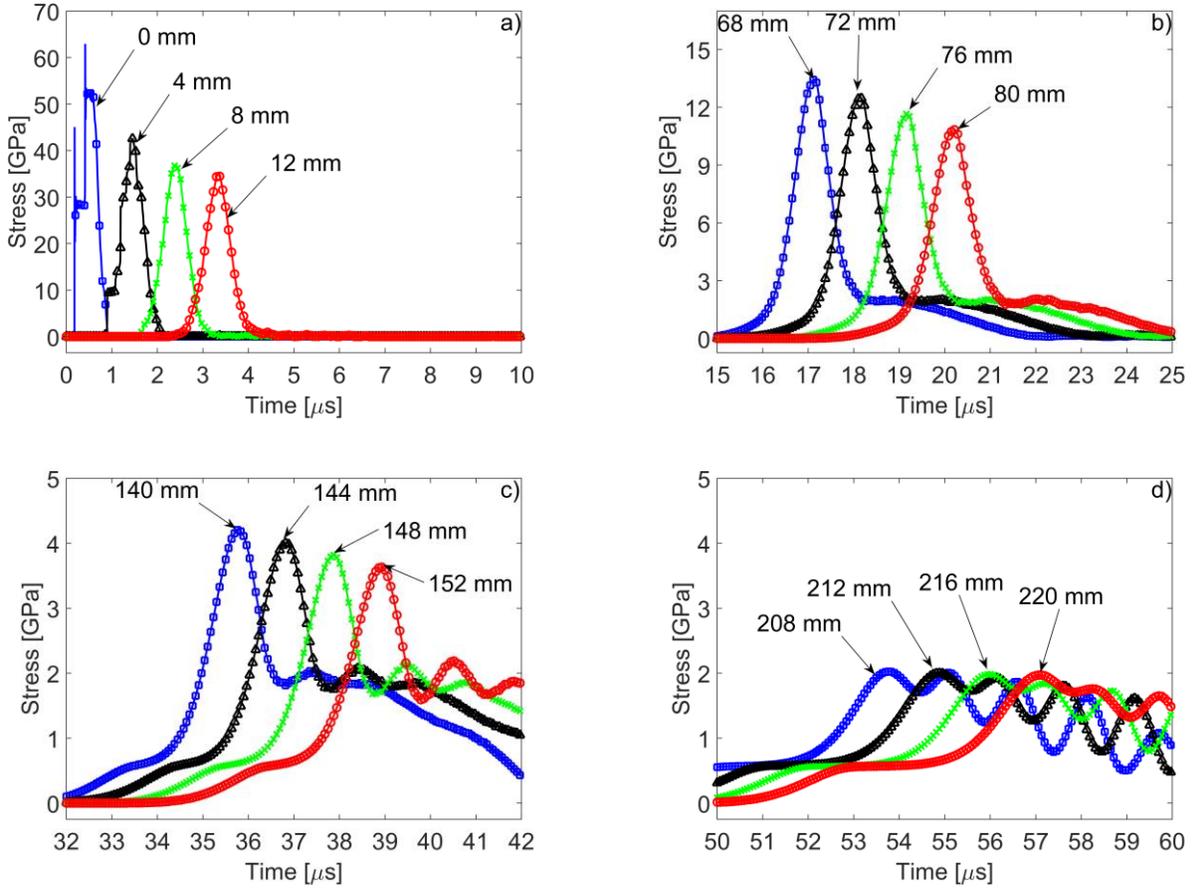

FIG. 12. (Color online) Stress pulse evolution in 1+1 laminate, data corresponds to the interfaces of Al/W layers at different depths (a) 0, 4, 8, and 12 mm (b) 68, 72, 76, and 80 mm (c) 140, 144, 148, and 152 mm (d) 208, 212, 216, and 220 mm. The pulse was generated by the impact of 2 mm Al plate at velocity of 2800 m/s.

The incoming pulse was quickly transformed into solitary like wave at depths 8 and 12 mm and at larger depths we observe leading solitary like wave followed by compression tail, if amplitude of the former was above 10 GPa. This resembles behavior of discrete strongly nonlinear granular chain [30, 31] except that strong nonlinearity resulted in decoupling of leading solitary wave from the compression wave, the latter being converted into shock wave. The difference between speed of a solitary like wave and compression wave with smaller amplitude in Al-W laminate is not as large as in strongly nonlinear granular chain preventing fast coupling of these waves before emerging of elastic precursor. We can expect a similar behavior if one of components in laminate exhibits a strongly nonlinear behavior.

**H. 0.5+0.5 Al/W Laminate, impact by 2 mm Al plate**

Laminates (2+2 and 1+1) impacted by 2 mm Al plate have shown the capability to form and propagate localized solitary like pulses (Fig. 10 and Fig. 12). It is interesting to investigate if 0.5+0.5 laminate will also support a single solitary like pulse or their train and traveled distance necessary for their formation when impacted by 2 mm Al flyer plate. In the case solitary wave is supported by this laminate, it is interesting to see its shape, characteristic space scale in relation to the cell size and rate of amplitude decay.

The ratio of characteristic duration of incoming pulse to laminate time scale introduced by mesostructure for this case is 2.5 which is identical to the case of 2+2 laminate impacted by 8 mm Al plate (Fig. 2). Therefore it is natural to expect the formation of a single solitary like wave with a small tail behind if mentioned time ratio is the main parameter determining the outcome of impact. Evolution of the generated pulse is shown in Fig. 13(a)-(d).

We can see that the incoming pulse was transformed into fast attenuating leading solitary like wave followed by compression wave at the distance about 4 mm away from the impacted end (Fig. 13(a)). It is interesting that the traveled distance required to form a solitary like pulse at the same characteristic duration of incoming pulse (thickness of impactor) is scaled with the cell size, compare shapes of pulses in 0.5+0.5 laminate at traveled distance 4 mm (Fig. 13(a)), in 1+1 laminate at a distance 8 mm (Fig. 12(a)) and in 2+2 laminate at distance 16 mm (Fig. 10(a)).

This behavior resembles the formation of two wave structure in discrete strongly nonlinear granular chain [30, 31] except that strongly nonlinear interaction between grains resulted in decoupling of leading solitary wave from the compression wave, the latter being converted into shock wave. The absence of separation of this waves in our case is probably due a relatively small difference between speed of a leading solitary like wave and following it compression wave with smaller amplitude in Al-W laminate in comparison with corresponding speeds in strongly nonlinear chain. We can expect a similar behavior if one of components in laminate exhibits a strongly nonlinear behavior.

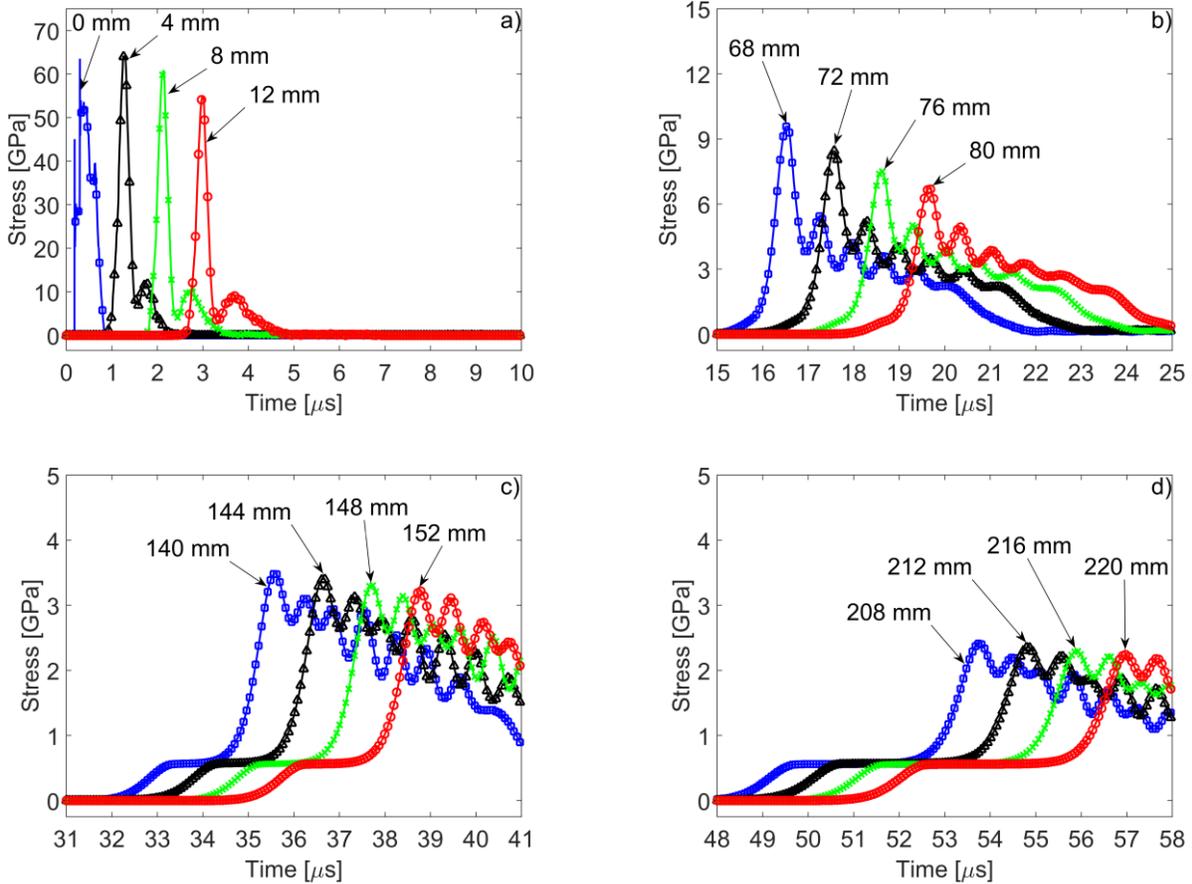

FIG. 13. (Color online) Stress pulse evolution in 0.5+0.5 laminate, data corresponds to the interfaces of Al/W layers at different depths (a) 0, 4, 8, and 12 mm (b) 68, 72, 76, and 80 mm

(c) 140, 144, 148, and 152 mm (d) 208, 212, 216, and 220 mm. The pulse was generated by the impact of 2 mm Al plate at velocity of 2800 m/s.

The strong attenuation of the leading pulse and probably dispersion prevents the formation of a train of localized stress pulses and create an attenuating oscillatory shock like profile (Fig. 13(b)-(d) already evident at distances about 68 mm. At very low stress levels (around 4 GPa) an elastic precursor dramatically changing the shape of propagating pulse can be observed (Fig. 13(c)-(d)), similar to previous case (Fig. 12(c)-(d)), although in case of 1+1 laminate, the elastic precursor appears closer to the impacted end. In this laminate, the scaling of the characteristics sizes of the localized pulse can be observed. At 68 – 72 mm depth, FWHM = 1.8 – 2.2 cell size and $\Delta = 2.1 - 2.2$ cell sizes.

## I. Comparison of solitary like wave shapes created by different initial conditions

It is interesting to compare solitary like waves with similar amplitudes in the same laminate resulting from different initial conditions caused by impactor with different thickness. If these waves generated by different initial conditions are quasistationary and similar than it may indicate that they are the result of balancing nonlinear and dispersive properties of material, as in the case with true solitary waves.

In Fig. 14(a)-(b) normalized profiles of solitary like waves with similar amplitudes propagating in the Al-W laminates (1+1 and 0.5+0.5) with different dissipative properties; excited by Al impactors with thickness 8 mm and 2 mm are presented. The profiles and durations of the corresponding incoming stress are quite different, for example the total durations of incoming pulses presented in Fig. 7(a) and Fig. 12(a) are 2.35 microseconds and 0.74 microseconds, correspondingly.

Four solitary like waves in normalized coordinates (distance divided by cell size) with very close stress amplitudes are shown in Fig. 14. Fig. 14(a) represents the wave profiles propagating in 1+1 and 0.5+0.5 laminates with real dissipative properties and similar maximum stresses close to 35 GPa, which were generated by the impact of Al flyer plates with thicknesses 2 mm and 8 mm. The respective travelling distances to reach similar amplitudes in these pulses are 40 mm and 104 mm (1+1 laminate) and 32 mm and 120 mm (0.5+0.5 laminate).

Wave profiles with the stress amplitude about 35 GPa are presented in Fig. 14(b) corresponding to the laminates with reduced dissipative properties (smaller $Y_{max}$), but with the same mesostructure as in Fig. 53(a). These profiles were formed at depths 182 mm (1+1 laminate), 90 mm (1+1 laminate), 60 mm (0.5+0.5 laminate), and 92 mm (0.5+0.5 laminate) correspondingly, where these quasistationary pulses had a similar stress amplitudes.

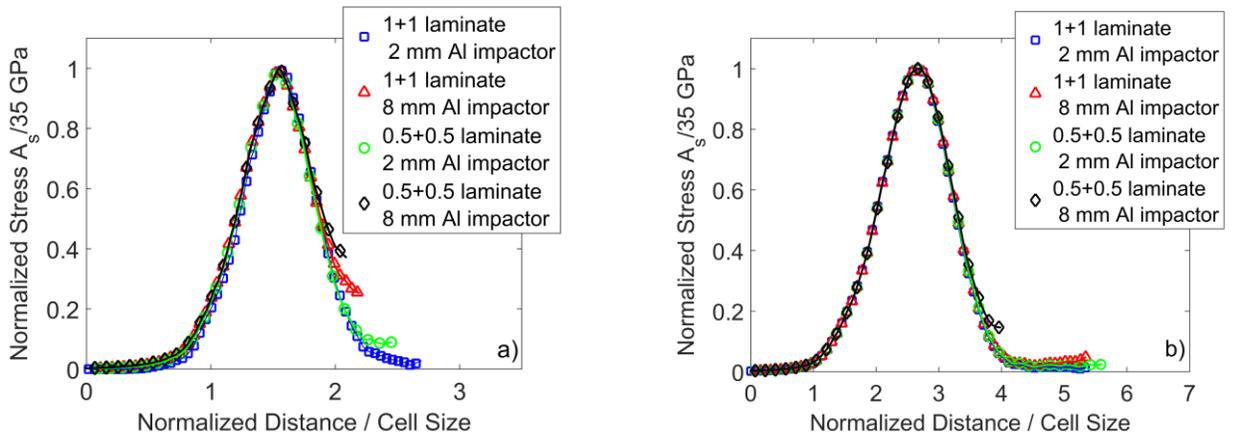

FIG. 14. (Color online) Comparison between the profiles of stress in solitary like waves with similar amplitudes in Al/W laminates with different cell sizes and dissipative properties generated by two different impactors with thickness 2 mm and 8 mm respectively. The stress is normalized by the following wave amplitudes: (a) laminates with real dissipative properties. 1+1 laminate, impacted by 2 mm Al plate, stress amplitude 34.9 GPa; 1+1 laminate, impacted by 8 mm Al plate, stress amplitude 34.8 GPa; 0.5+0.5 laminate, impacted by 2 mm Al plate, stress amplitude 35.2 GPa; and 0.5+0.5 laminate, impacted by 8 mm Al plate, stress amplitude 35 GPa. (b) laminates with artificially low $Y_{max}$. 1+1 laminate, impacted by 2 mm Al plate, stress amplitude 35.2 GPa; 1+1 laminate, impacted by 8 mm Al plate, stress amplitude 35.9 GPa; 0.5+0.5 laminate, impacted by 2 mm Al plate, stress amplitude 35 GPa; and 0.5+0.5 laminate, impacted by 8 mm Al plate, stress amplitude 35 GPa.

The comparison between profiles of these waves demonstrates that they are similar in normalized coordinates despite difference in their dissipative properties. For the case of the laminate with real properties, the FWHM = 1.3 and for the material with small $Y_{max}$, FWHM = 1.3. This comparison confirms that these waves indeed are a result of the material properties (balancing dispersion caused by the periodic mesostructure and nonlinearity) and do not depend on the initial conditions that generated them, similar to properties of true solitary waves.

### J. Head on collision of solitary like waves

One of the main properties of true solitary waves is that they preserve their shapes after collision [9, 16]. It is interesting if solitary like waves described in previous section behave in a similar way. To investigate collision of solitary like waves the 2+2 laminates with real material properties was impacted on both ends as shown in Fig. 15. The length was $L_c$ = 202 mm. The impactors thickness was $L_i$ = 6 mm. They had velocities $V_a$ = 3500 m/s and $V_b$ = 3200 m/s

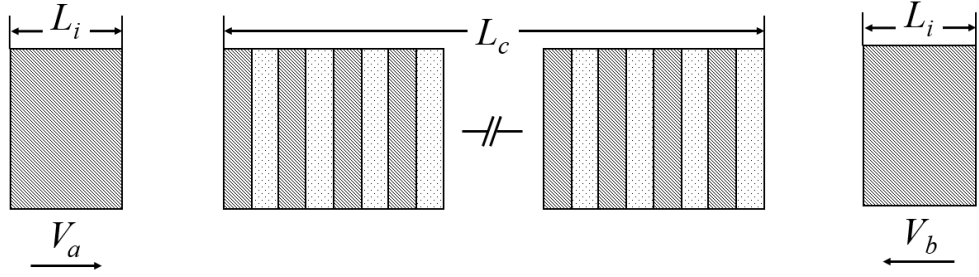

FIG. 15. Laminate impacted at both ends to investigate collision of solitary like waves.

Fig. 16 (a)-(b) show both localized waves before collision at 60 mm from the impacted ends. It is clear that they have almost symmetrical shape with small amplitude tails. The difference in their amplitude is caused by different velocities of impactors on corresponding ends.

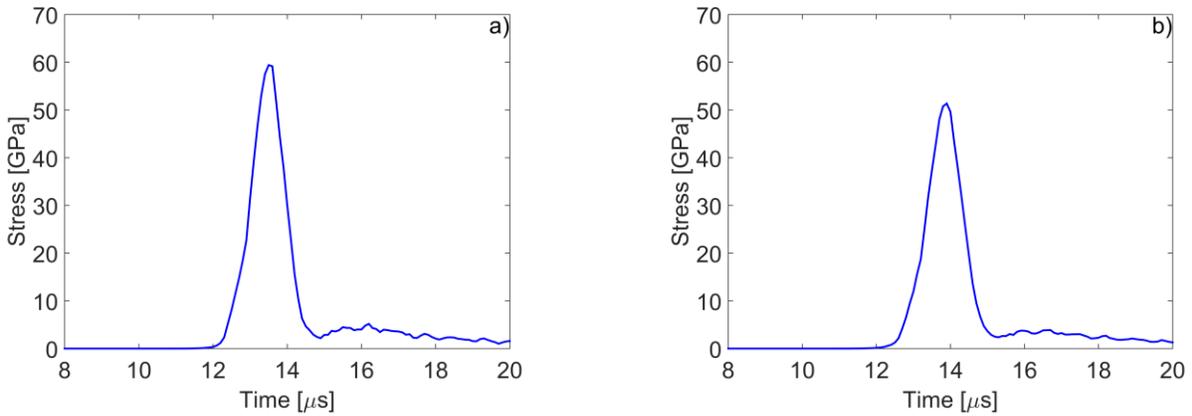

FIG. 16. (a) Traveling wave at 60 mm depth from the left end of the laminate with real properties (b) Traveling wave at 60 mm depth from the right end of the laminate. Both waves are shown before collision.

Fig. 17 corresponds to the depths 101 mm from both ends of laminate with real properties where the two waves propagating in the opposite directions meet. The amplitude of the resulting pulse is increased as well as the amplitude of the tail, its duration was close to the durations of pulses before collision.

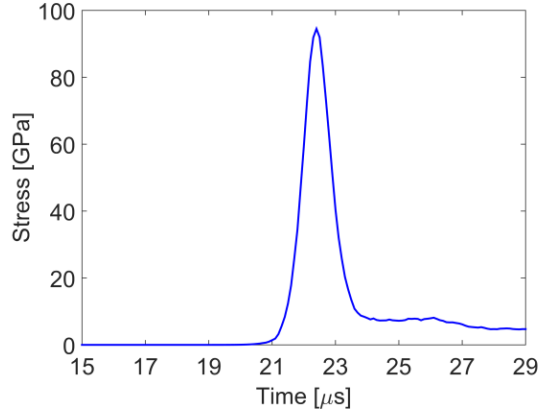

FIG. 17. Resulting pulse due to collision of two waves propagating in the opposite direction at the middle of the laminate (at 101 mm depths from both ends).

Pulses after their collision are shown in Fig. 18 (a)-(b) at distances 142 mm from corresponding ends. We can see that after collision we have two localized waves with different amplitudes, which are smaller than amplitudes before collision (Fig. 16), mostly because waves decay by travelling additional distances. It should be mentioned that the amplitudes of tails increased after collision due to dissipation.

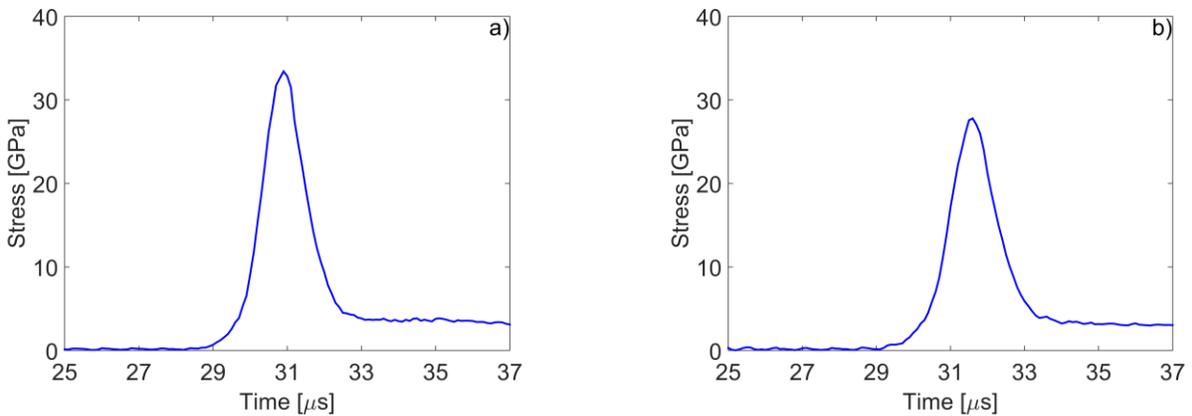

FIG. 18. (a) Traveling wave at 142 mm depth from the left end of the laminate with real properties (b) Traveling wave at 142 mm depth from the right end of the laminate. Both waves are shown after collision.

Collision of true solitary waves result in the phase shift [9, 16]. To investigate if the phase shift is also characteristic for the collision of solitary like waves, we compare two initially identical solitary like waves created by the impact on the left end (shown in Fig. 16(a)), but traveling the same distance without collision. Fig. 19(a) presents a comparison of the same wave travelled without collision and after head on collision at 122 mm from the left end. These two waves are superposed in Fig. 19(b) to demonstrate their similar profiles.

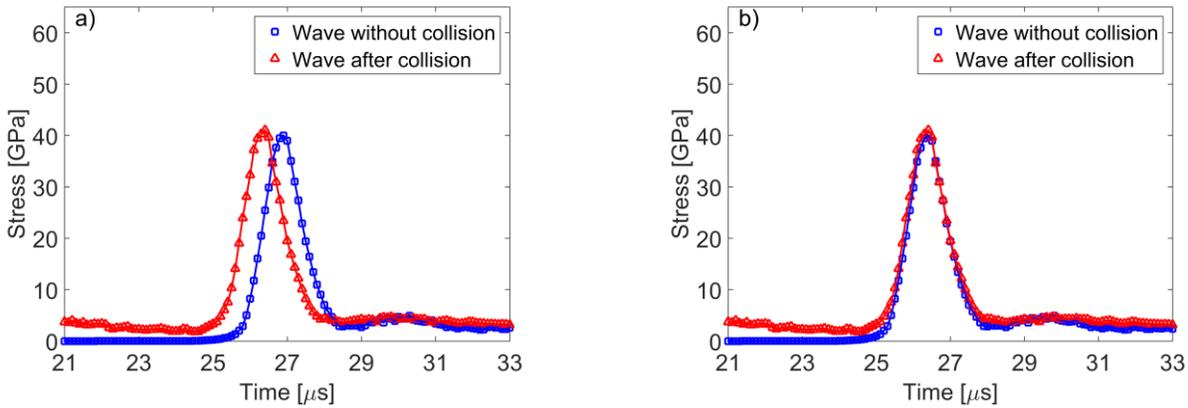

FIG. 19. (Color Online) (a) The phase shift between two initially identical solitary like waves created by the impact on the left end traveling the same distance 122 mm without collision and after head on collision; (b) the same waves are superposed to demonstrate their similar profiles.

From Fig. 19(a) it is clear that a phase shift has occurred after collision, similar to phase shift observed after collision of true solitary waves. The two superposed waves in Fig. 19(b) demonstrate that they are very similar resembling the behavior of classic solitons, which completely reconstruct their shapes after collision. The difference between these solitary like waves and true solitary waves is caused by the decay of the former due to the significant influence of dissipation at this level of stresses.

## IV. THEORETICAL APPROACH

The considered laminate material has a periodic structure and also exhibits significantly nonlinear behavior mostly due to nonlinearity in constitutive equation. Combination of these properties support the propagation of solitary like waves under certain conditions of dynamic loading as observed numerically in previous sections. Papers [16, 17, 20, 21, 22] introduce long wave approximation resulting in Boussinesq like wave equation to describe propagation of nondissipative solitary like wave in laminates with relatively small dynamic elastic strains, about $10^{-2}$. Nonlinearity in these papers was introduced based on nonlinear elasticity.

In this section we will combine dispersive and nonlinear properties of laminate incorporating them into KdV type equation, which support solitary waves, and compare them with results of numerical calculations of discrete system. Of course this approach, especially when resulted in a stress pulse with a dimensions comparable to the cell size of laminate, needs verification by numerical calculations of real discrete system. But if successful, it provides the scaling dependence of parameters of localized solitary like stress pulse of high amplitude on physical and geometrical parameters of laminates.

The dispersion relation for laminated materials can be found in [28], this expression considers the multiple reflections at the interfaces in the laminated material.

$$\cos(kd) = \cos\left(\frac{w d_a}{C_a}\right) \cos\left(\frac{w d_b}{C_b}\right) - \frac{1}{2}\left(\frac{Z_a}{Z_b} + \frac{Z_b}{Z_a}\right) \sin\left(\frac{w d_a}{C_a}\right) \sin\left(\frac{w d_b}{C_b}\right), \quad (11)$$

where $k$ is the wave number, $w$ is wavelength, $d_a$ and $d_b$ are the respective sizes of each layer, cell size $d = d_a + d_b$, $C_a$ and $C_b$ refer to the respective sound speed in each layer and $Z_a$ and $Z_b$ represent the respective impedances for each layer ($Z_i = \rho_i C_i$). In the limit of long wave approximation, $\lambda \gg d$ we get the following dispersion relation

$$w^2 = \frac{k^2 d^2}{\left[\frac{d_a^2}{C_a^2} + \frac{d_b^2}{C_b^2} + \left(\frac{Z_a}{Z_b} + \frac{Z_b}{Z_a}\right)\left(\frac{d_a d_b}{C_a C_b}\right)\right]}\left[1 - \frac{k^2 d^2}{12}\right] = C_0^2 k^2 \left(1 - \frac{k^2 d^2}{12}\right). \tag{12}$$

The first term in the Eq. 12 provides an explicit expression for $C_o$ for the laminate material, which is equivalent to the averaging approach presented on [16]

$$C_o^2 = \frac{d^2}{\frac{d_a^2}{C_a^2} + \frac{d_b^2}{C_b^2} + \left(\frac{Z_a}{Z_b} + \frac{Z_b}{Z_a}\right)\left(\frac{d_a d_b}{C_a C_b}\right)}. \tag{13}$$

In our numerical calculations, which included dissipation and more than order of magnitude larger strains than in [16, 17, 20, 21, 22] (about 10$^{-2}$) we also observed solitary like, slowly attenuating localized waves. But at these much higher strains it is more appropriate to introduce nonlinearity using Hugoniot relation. Hugoniot parameters naturally reflect the nonlinear properties of materials because shock wave is a typical example of the phenomena supported by combination of nonlinear and dissipative properties. It should be mentioned that temperatures in solitary like waves are smaller compare to shock temperatures at similar stress amplitudes. But because input of temperature to the stress at given specific volume is relatively small the use of Hugoniot curve is appropriate. To introduce nonlinearity in the constitutive equation for each of the materials in laminate we consider Hugoniot curves (Eq. 14) in stress versus strains coordinates for Al and W and their corresponding approximations using second powers of strains (Eq. 18). The stress along the Hugoniot is defined as:

$$P = \frac{C_o^2 (V_o - V)}{[V_o - s(V_o - V)]^2}, \tag{14}$$

where $V_o$ and $V$ correspond to the specific volume of the material at initial and deformed configuration respectively and $s$ refers to the first coefficient of the Hugoniot curve on the D-u plane (Shock-Particle Velocity relation) which has the form

$$D = C_o + s\,u. \tag{15}$$

In this equation a single coefficient $s$ is representing nonlinear behavior of material in shock wave conditions. Strain in shock wave in terms of specific volume can be written as:

$$\epsilon = \frac{V_o - V}{V_o}. \tag{16}$$

By combining Eqn. 14 and 16 we obtain Hugoniot stress strain relation

$$P = \frac{C_o^2}{V_o}\left[\frac{\epsilon}{(1-s\,\epsilon)^2}\right]. \tag{17}$$

At relatively low values of strains (below 0.15 for Al and 0.1 for W) we can approximate the behavior of Al and W using the following equation with corresponding coefficients for each material:

$$P = \frac{C_o^2}{V_o}[\epsilon(1 + 2s\epsilon)]. \tag{18}$$

The Hugoniot curves and their approximations by Eq.18 are presented in the Figs. 20(a) and 20(b). For the comparison, zero Kelvin compression curves, isentrope (Al [29], W estimated from [33]) and curves based on third order elastic constants, are also presented. The latter ones are described by the following equation

$$\sigma = \gamma\epsilon + \frac{\beta}{2}\epsilon^2, \tag{19}$$

where coefficients $\gamma$ and $\beta$ are defined by:

$$\gamma = \lambda + 2\mu, \tag{20}$$

$$\beta = 3(\lambda + 2\mu) + 2(A + 3B + C), \tag{21}$$

in which λ and μ are the Lamé parameters and *A, B* and *C* correspond to the 3rd order elastic constants [34].

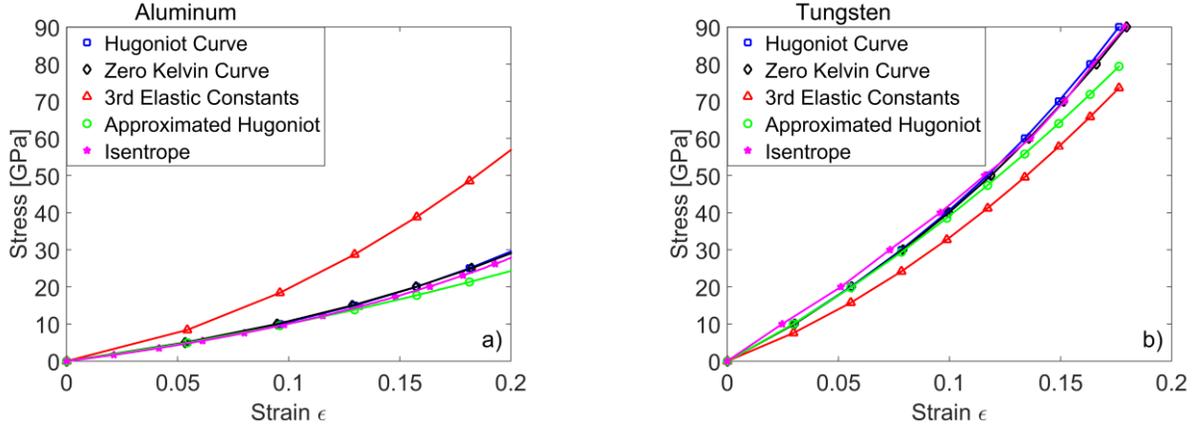

FIG. 20. (Color online) Nonlinear stress-strain relations for Al (a), and W (b) based on Hugoniot curves, static compression based on third order elastic constants, zero Kelvin curves, Isotherm and approximation of Hugoniot curves using terms only with second power law of strains, Eqn. 16.

It is clear that using third order elastic constants in our range of stresses is not adequate. At the same time deviation of stresses on Hugoniot curve from the values corresponding to zero Kelvin curve at the same strains are within a few percent for Al and W at stresses below 40 GPa (strains below 0.25 for Al and below 0.1 for W). Thus we can use Hugoniot curves approximated by Eq. 18 as a reasonable representation of isentropic behavior. It is appropriate taking into account that on the steady state of propagation of localized stress pulses materials experience isentropic compression paths, which are between the zero Kelvin curve, and Hugoniot states. It should be emphasized that we are looking for a non-dissipative description of the single stress pulse despite existing dissipation at this level of stresses. Results of numerical calculations presented above demonstrates that dissipation, resulting in the pulse attenuation, does not change the nature of the pulse coming from balance of nonlinearity and dispersion similar to the case of strongly nonlinear solitary waves in granular lattices [9].

To find the nonlinearity coefficient in stress strain relation for the laminate material we consider its single cell with unit area as depicted on Fig. 21.

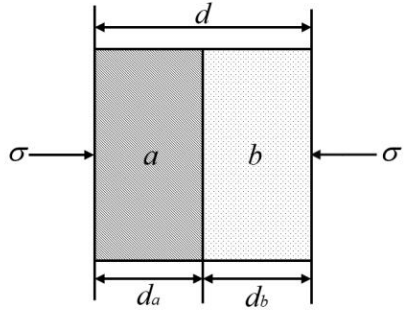

FIG. 21. Single cell with unit area made of material *a* and *b*.

We need to find a constants $K_{eq}, \alpha_{eq}$ describing the behavior of the unit cell under compression with effective total strain $\varepsilon_t$. We define the total strain of the deformed cell as:

$$\epsilon_t = \frac{(d_{Al}+d_W)-(d_{Al_o}+d_{W_o})}{d_{Al_o}+d_{W_o}} = \frac{d_{Al}-d_{Al_o}}{d_{Al_o}+d_{W_o}} + \frac{d_W-d_{W_o}}{d_{Al_o}+d_{W_o}} = \epsilon_a \left(\frac{d_{Al_o}}{d_{Al_o}+d_{W_o}}\right) + \epsilon_b \left(\frac{d_{W_o}}{d_{Al_o}+d_{W_o}}\right) = \epsilon_{Al}\,\tau +$$

$$\epsilon_W(1-\tau)\,, \tag{22}$$

where $d_{Al_o}, d_{W_o}$ represent the original length of Al, W and the symbols $d_{Al}, d_W$ are related to a deformed cell. For the system being in equilibrium, $\sigma_{Al} = \sigma_W = \sigma_t$, where is the force applied to the cell. The equation for a total strain including a nonlinear term quadratic with respect to $\epsilon^2$ is represented by the following expression

$$\sigma_t \approx \frac{K_{Al}K_W}{(1-\tau)K_{Al}+\tau K_W}\epsilon + \frac{(\tau K_W^3 \alpha_{Al}+(1-\tau)K_{Al}^3 \alpha_W)}{((1-\tau)K_{Al}+\tau K_W)^3}\epsilon^2, \tag{23}$$

This gives us the expressions for coefficient of nonlinearity $\alpha_{eq}$ and linear elastic modulus $K_{eq}$ representing the global response of the cell

$$\alpha_{eq} = \frac{(\tau K_W^3 \alpha_{Al} + (1-\tau) K_{Al}^3 \alpha_W)}{((1-\tau) K_{Al} + \tau K_W)^3}, \tag{24}$$

$$K_{eq} = \frac{K_{Al} K_W}{(1-\tau) K_{Al} + \tau K_W}. \tag{25}$$

Apparently that coefficient of nonlinearity is the same for laminates with the same volume of components. The wave equation being a long wave approximation of discrete, non-dissipative system and having the same dispersive relation and reflecting nonlinear behavior is Boussinesq equation

$$U_{tt} = C_o^2 U_x + \beta U_{xxxx} - \psi U_x U_{xx}, \tag{26}$$

where coefficients $\beta$ and $\psi$ are related to materials parameters in the following way:

$$\beta = \frac{d^2 C_o^2}{12}. \tag{27}$$

Unlike the coefficient of nonlinearity and sound speed which is the same for laminates with the same volume ratio of components coefficient of dispersion depends on the characteristic scale of laminate $d$. Parameter $\psi$ is the coefficient related to the nonlinear part of the force applied to the cell composed from the elements of the discrete system

$$\psi = \frac{2\alpha C_o^2}{K_{eq}} \tag{28}$$

This equation can be converted to KdV equation in the same approximation:

$$\zeta_t + C_o \zeta_x + S \zeta_{xxx} + \nu \zeta \zeta_x = 0 \tag{29}$$

$$S = \frac{\beta}{2}, \tag{30}$$

$$\nu = \frac{\psi}{2 C_o}, \tag{31}$$

KdV equation has solitary wave solution of the form:

$$\zeta = \zeta_m \operatorname{sech}^2\left(\left(\frac{\psi\zeta_m}{12\beta}\right)^{\frac{1}{2}}(x-vt)\right). \tag{32}$$

The equation for the full-width at half-maximum (FWHM) of this pulse ($w$) expressed in terms of the cell size, thicknesses of layers and linear and nonlinear properties of components as well as an amplitude of stress pulse is presented below

$$FWHM = \frac{1.76 d K_{eq}^{1/2}}{(2\alpha_{eq}\zeta_m)^{1/2}} = \frac{1.76 d^{1/2}}{2\zeta_m^{1/2}}\left[\frac{(C_{Al}^2\rho_{Al}d_W+C_W^2\rho_W d_{Al})^2}{(C_W^4\rho_W^2 S_{Al}d_{Al}+C_{Al}^4\rho_{Al}^2 S_W d_W)}\right]^{1/2}. \tag{33}$$

It is interesting that though the FWHM of the pulse in laminate $w$ is scaled linearly with the cell size $d$ the individual thicknesses of layers are also affecting the size of the solitary wave being present in values of $K_{eq}$ and $\alpha_{eq}$. It means that laminates with the same cell size $d$ will have different values of $w$ if the thicknesses of individual layers are different.

The speed of the solitary wave is given by

$$V = C_o + \frac{v}{3}\zeta_m, \tag{34}$$

$$v = \frac{\psi}{2C_o} = \frac{2dC_{Al}^2\rho_{Al}C_W^2\rho_W(C_{Al}^6\rho_{Al}^3 K_W S_W d_W + C_W^6\rho_W^3 K_{Al} S_{Al} d_{Al})}{K_{Al}K_W(C_{Al}^2\rho_{Al}d_W+C_W^2\rho_W d_{Al})(\rho_{Al}d_{Al}+d_W\rho_W)} =$$

$$\frac{2C_{Al}^2\rho_{Al}C_W^2\rho_W(C_{Al}^6\rho_{Al}^3 K_W S_W d_W/d + C_W^6\rho_W^3 K_{Al} S_{Al} d_{Al}/d)}{K_{Al}K_W(C_{Al}^2\rho_{Al}d_W/d+C_W^2\rho_W d_{Al}/d)(\rho_{Al}d_{Al}/d+\rho_W d_W/d)}. \tag{35}$$

Now we explore if this approximation satisfactory describes the shape of solitary like stress pulses and their speeds observed in numerical calculations. Figs. 22 (a)-(d) presents the shapes of stress pulse found in LS-DYNA numerical simulations and KdV solitary wave solution (Eq. 32) for maximum stresses below 35 GPa where cold curve and isentropic compression are very well approximated by nonlinear equation (Eq. 18). The pulse shapes generated in the numerical calculations performed with real material properties as well with the one with artificially small $Y_{max}$ are shown in Fig. 22.

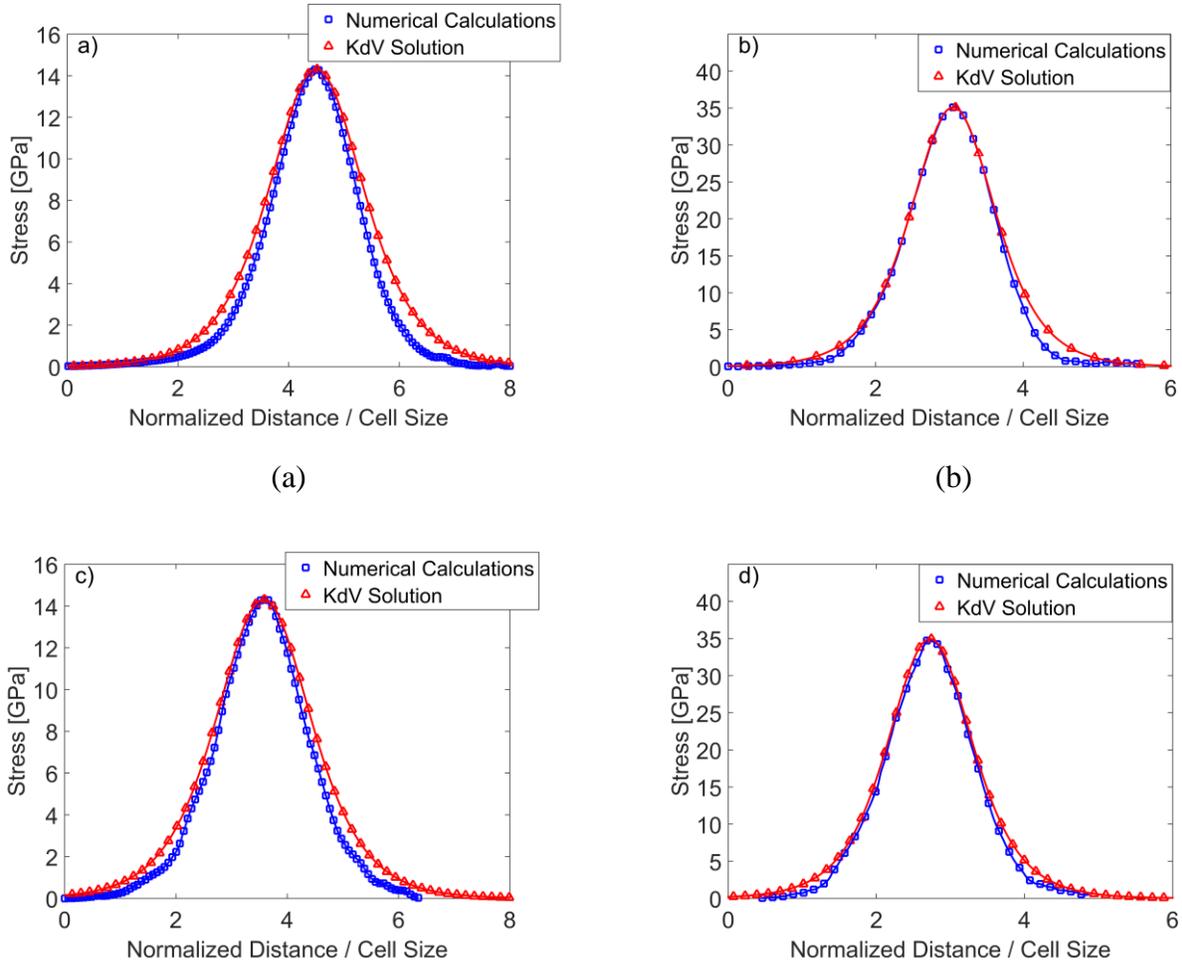

FIG. 22. Comparison of KdV solitary solution with nonlinearity parameters taken from Hugoniot curve (Eq.18) to the shapes of solitary like waves found on numerical simulations: (a) The wave in numerical simulation corresponding to the depth 72 mm in the 2+2 laminate with artificially small $Y_{max}$ laminate impacted by a 2 mm Al plate with velocity 2800 m/s. (b) The wave in numerical simulation corresponding to the depth 32 mm in the 1+1 laminate with artificially small $Y_{max}$ impacted by a 2 mm Al plate with velocity 2800 m/s. (c) and (d) comparison of waves in corresponding laminates (2+2 and 1+1) with real properties.

We can see that space scales and shapes of stress pulses observed in numerical calculations are satisfactory described by long wave KdV solitary wave solution despite that characteristic width of pulse being close to the cell size and dissipation present in numerical calculations. It should be mentioned that the full-width at half-maximum (FWHM) for these pulses is only about 2 cells. The pulse in numerical calculation is narrower than KdV solution, the FWHM of the KdV solitary wave solution is about 12% larger than corresponding pulse width in numerical calculations in case of solitary like wave in 2+2 laminate (Fig. 22(c)).

It is interesting to check if a presented solitary wave solution is also a reasonable approximation for a broader range of wave amplitudes, specifically for significantly larger stresses where nonlinear behavior deviates more significantly from approximation described by Eq. 18. The corresponding shape of the stress pulse at these stresses is shown in Fig. 23.

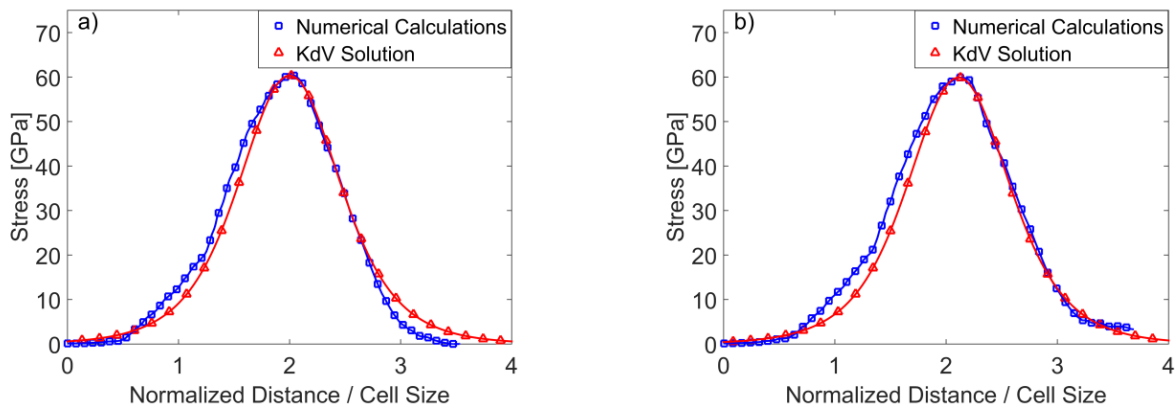

FIG. 23. (Color online) Comparison of KdV solitary solution with nonlinearity parameters taken from approximated Hugoniot curve (Eq.16) to the wave found on numerical simulations. The wave in numerical simulation corresponds to the depth 132 mm in a 2+2 laminate impacted by 8 mm Al plate with velocity 2800 m/s and artificially small Ymax (a) and in laminate with real material properties (b).

The value of FWHM in this case is about one cell size only. It is amazing that KdV solitary wave, obtained as a solution of a wave equation being a weakly nonlinear and long wave approximation of discrete system, is still a satisfactory approximation for the shape of the very short localized pulse. Thus Equation 33 provides a correct scaling of the size of solitary like pulses and a reasonable description of their shapes in a very broad range of stress amplitudes.

Another property of KdV solitary waves is a linear dependence of their speed on the stress amplitude (Eq. 36). Fig. 24 presents a comparison of the dependence of speed of pulses on the stress amplitude found in numerical calculations for laminates with different cell sizes with the similar dependence for KdV solitary wave (Eq. 36).

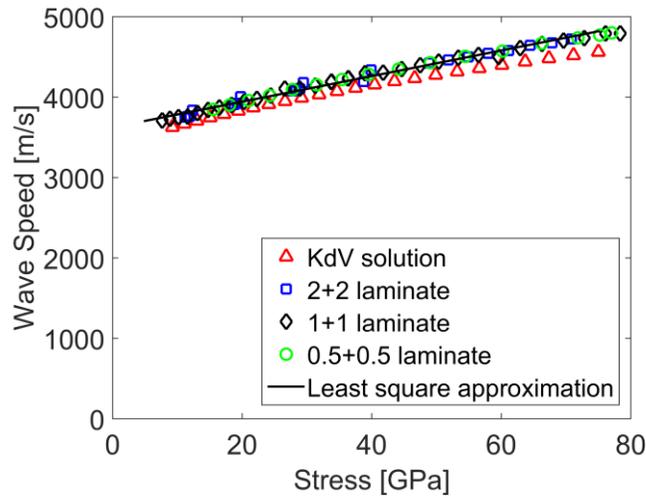

FIG. 24. Dependence of the speed of the localized wave in Al-W 2+2, 1+1, and 0.5+0.5 laminates found in numerical calculations on maximum stress (on the interface between Al and W) and corresponding Eq. 33 for the speed of KdV solitary wave.

It is clear that linear relationship between speed of the pulse and maximum stress found for KDV solitary wave is a satisfactory approximation in whole investigated range of stress amplitudes for different laminates. It is important that the speed of the solitary like wave in numerical calculations is not dependent on the cell size, as can be expected from the proposed analogy with KdV soliton. It is interesting that at the same cell size, but at different individual thicknesses of layers (determining the relative volume of component) the speed of solitary wave will be different. This dependence is represented by values of $K_{eq}$ and $\alpha_{eq}$ depending on volume fraction of components. It means that laminates with the same cell size $d$ will have different values of the slope of the speed versus stress amplitude if the thicknesses of individual layers are different.

It should be mentioned that sound speed for the laminate, based on Eq. 13 is $C_{eq} = 3381$ m/s. From the graph presented in Fig. 24 we can see that speed of the solitary wave in KdV approximation and also in numerical calculations is larger than sound speed in the laminate in a long wave approximation, thus this pulse is supersonic with respect to the long wave sound speed in the laminate. But its speed in the all investigated range of stresses is smaller than sound speed in Al and at maximum stresses below 30 GPa is also smaller than in W. As a result a short wave length disturbances with the space scale compared to the thickness of layers can escape into the area in front of propagating pulse leaking the energy from the pulse and contributing to the amplitude attenuation. This mechanism is additional to the dissipation and presents another reason that these propagating pulses are not truly solitary waves despite being satisfactory described by KdV solutions. This mechanism also contributes to the nonelastic collision of these localized pulses.

## IV. CONCLUSIONS

The existence of solitary-like localized waves in laminate material Al-W in a broad range of stress amplitude was demonstrated using numerical calculations. It was shown that the dissipation due to viscoplastic behavior causes significant decay of amplitude of these waves, but their space scale and shape are closely approximated by KdV solution.

These solitary-like waves exhibit behavior similar to classical KdV solitons, e.g., a dependence of speed and width on the stress amplitude. The different durations of incoming pulse, with respect to characteristic time scale of laminate, result in either the formation of only one localized solitary like wave, a train of such waves or in oscillatory shock like wave. It has also been shown that interaction between this waves result in a phase shift, which is similar to behavior on classic solitons, although it also shows a nonelastic behavior.

A theoretical framework based on a weekly non-linear approach and resulting KdV equation supporting solitary wave was presented. The approximation made in this paper rely on readily available shock Hugoniot data used to introduce nonlinearity parameter and on the exact dispersion relation for linear elastic laminate. The shape and speed of the localized waves found in numerical calculations are in a satisfactory agreement with the KdV solitary wave in a broad range of stress amplitudes.

This approach allowed to arrive to the analytic equation for width and speed of the observed localized pulse using nonlinear properties and geometry of laminate. It allows to design layered materials for favorable for protection, e.g., to prevent spall and understand the nature of pulses propagating under extremely short pulse loading, e.g., produced by powerful lasers.